\documentclass[aps,twocolumn,10pt, nofootinbib]{revtex4-2}
\pdfoutput=1

\usepackage{amsfonts}
\usepackage{amsmath,amssymb,amsthm}
\usepackage{commath}
\usepackage{bm}
\usepackage{dcolumn}
\usepackage{epsfig}
\usepackage[T1]{fontenc}
\usepackage{graphicx}
\usepackage{graphics}
\usepackage[pdftex]{hyperref}
\usepackage{hhline}
\usepackage[utf8]{inputenc}
\usepackage{latexsym}
\usepackage{multirow}
\usepackage[normalem]{ulem}
\usepackage{rotating}
\usepackage{color}
\usepackage{xcolor}
\usepackage{tikz}
\usetikzlibrary{positioning, arrows.meta, fit,decorations.pathreplacing,calc}

\newcommand\be{\begin{equation}}
\newcommand\ba{\begin{eqnarray}}
\newcommand\ee{\end{equation}}
\newcommand\ea{\end{eqnarray}}

\begin{document}

%----- Title -----%
\hspace{15 cm}
\title{The Shape of Eccentricity: \\ Rapid Classification of Eccentric Binaries with the Wavelet Scattering Transform}

\author{Priscilla Canizares}
\email{pc464@cam.ac.uk}
\affiliation{DAMTP, Centre for Mathematical Sciences, University of Cambridge, Wilberforce Road, Cambridge CB3 0WA, UK}

\author{Seppe J. Staelens}
\email{ss3033@cam.ac.uk}
\affiliation{DAMTP, Centre for Mathematical Sciences, University of Cambridge, Wilberforce Road, Cambridge CB3 0WA, UK}
\affiliation{Leuven Gravity Institute, KU Leuven,
Celestijnenlaan 200D box 2415, 3001 Leuven, Belgium}

\author{Isobel Romero-Shaw}
\email{romero-shawi@cardiff.ac.uk}
\affiliation{Gravity Exploration Institute, School of Physics and Astronomy, Cardiff University, Cardiff, CF24 3AA, UK}

\begin{abstract}
The gravitational-wave (GW) detections reported by the LIGO--Virgo--KAGRA (LVK) Collaboration have so far been consistent with quasi-circular compact binary coalescences. However, a small fraction of binaries driven to merge through dynamical interactions in dense stellar environments or in field triples may retain measurable orbital eccentricity upon entering the sensitive frequency band of the LVK detectors. A confident measurement of eccentricity in the LVK band would constitute decisive evidence for such dynamically driven mergers. Eccentric waveform models, however, are computationally expensive, so performing production-level inference on \textit{all} detected signals is an inefficient use of resources when eccentric signals are expected to be rare. In this work, we propose a parsimonious strategy that analyses the \textit{wavelet scattering transform} (WST) coefficients of GW signals as an intermediate step between \textit{detection} and \textit{analysis}: each signal is assessed for the potential presence of eccentricity in order to provide rapid recommendations for which signals should undergo production-level eccentric inference. By compressing the signals and working with lower-dimensional classifiers, this approach reduces computational complexity and thereby speeds up classification. We evaluate the discriminatory power of the WST using a simple one-dimensional convolutional neural network trained on a large set of synthetic waveforms injected into realistic noise. We find that the WST representation enables effective discrimination between eccentric and quasi-circular binaries; we discuss the advantages of its compact, multi-scale representation and benchmark it against both simpler and more complex classifiers. Using an effective one-body waveform model, our pipeline achieves a detection accuracy of $\sim$$64\%$ at a false-alarm rate of $10\%$, with an AUC of $0.844$ and an average precision of $0.876$. Finally, we investigate the WST coefficients' ability to distinguish eccentricity from spin-induced precession, and find that it performs well across a range of spin-precession magnitudes. We conclude that the WST is a powerful tool for identifying the fingerprints of eccentricity in compact binary coalescence GW signals.
\end{abstract}

%----- Date and Preprint Number -----%
\date{\today}

%\preprint{}

%----- MAKETITLE -----%
\maketitle

\hspace{10pt}

\normalsize

\section{Introduction}
\label{sec:intro}
Gravitational wave (GW) detections made by the LIGO-Virgo-KAGRA (LVK) Collaboration are yielding exciting inferences about the production mechanisms of merging compact binaries, thus providing better constraints on their formation environments. The most recent Gravitational Wave Transient Catalog, GWTC-5 \citep{LIGOScientific:2026wfs}, contains 267 candidates with false alarm rate (FAR) $<1$~yr$^{-1}$ \citep{LIGOScientific:2026ctl}; the properties of this population appear consistent with being produced via a variety of formation mechanisms, including isolated field triples \citep[e.g,][]{Stegmann25} and dynamical formation in dense star clusters \citep[e.g.,][]{2025arXiv250904637A, 2025arXiv251105316T}. In contrast to fully isolated binary evolution---which is expected to produce only quasi-circular mergers in the LVK frequency range \citep{P_64}---a small fraction of dynamically-formed and tertiary-influenced mergers should retain measurable eccentricity close to merger \citep[e.g.,][]{Antonini17, Samsing18}.

LVK catalogues have not included eccentricity measurements so far: historically, development of quasi-circular waveform models was prioritised for detection and analysis of the first detections, leading to inefficient eccentric waveform models. Recently, however, short-author works have used increasingly efficient new eccentric waveform models to place constraints and, in some cases, measure non-negligible eccentricity in CBC signals \citep[e.g.,][]{IRS22, Gupte24, Planas25, Morras25, Xu25, Paul_2025, Phukon_2025}. None of the eccentric candidates are confident detections of eccentricity: their interpretation is clouded by missing merger or spin physics in the waveform models, possible confusion with spin-induced precession \citep{Romero_Shaw_2023, Divyajyoti24, Divyajyoti25, Tibrewal26}, data quality issues \citep[e.g.,][]{2022PhRvD.106j4017P}, or low signal-to-noise ratio (SNR) \citep[e.g.,][]{2025PhRvD.112f3052R}. Nonetheless, with growing evidence for dynamical and/or triple-origin mergers in the data, we could anticipate confident detections of eccentricity in the near future \citep{2021ApJ...921L..43Z}: such formation scenarios have a predicted eccentric merger output of $\sim10\%$.

There are, however, some hard-to-quantify barriers to these expectations. LVK matched-filter searches are tuned to quasi-circular signals, which significantly reduces their sensitivity to eccentric CBC signals \citep[e.g.,][]{2010PhRvD..81b4007B, 2021ApJ...921L..43Z, 2024PhRvD.110d4013G}. While unmodelled searches may capture some additional high-mass eccentric sources, we remain observationally biased against observing inspiral-dominated eccentric signals \citep{2020PhRvD.102d3005R, 2024PhRvD.110d4013G, 2025arXiv251216289R}. Until recently, eccentric waveform models were near-prohibitively expensive, requiring novel methods like importance-sampling with likelihood reweighting \citep{2019MNRAS.490.5210R} or machine learning \citep{Gupte24} to make their use in analysis feasible. More efficient eccentric waveform models now exist \citep[e.g,][]{Morras, SEOBNRv5EHM, 2025PhRvD.112l1503A, ravichandran2024}, but they retain the limitations described above, and/or are still much slower than their quasi-circular counterparts. Exacerbating this expense, the LVK's detection efficiency has increased impressively since the first GW detection, and there are now hundreds of new compact binary coalescence (CBC) signals per observing run. With eccentric mergers constituting a relatively small fraction of the underlying physical population, and with search methods biased against detecting them, we argue that running eccentric analyses on all LVK candidates may not be well-justified.

It is reasonable, therefore, to propose a computationally efficient selection step in-between detection and eccentric analysis. Machine learning is optimally suited to such a problem, being highly successful in areas requiring pattern recognition and prioritising speed. Indeed, \citep{ravichandran2024} designed and trained a two-dimensional convolutional neural network (2D-CNN) to study Q-scans of GW data and identify the presence of eccentricity in the signal. Similarly, \cite{Khalouei_2025} proposed a transformer-based approach to detect eccentricity from GW scalograms.

Here, we address this problem using a new approach: employing the wavelet scattering transform (WST) to represent GW strain data. This circumvents the need to process scans or images by operating directly on a compressed representation of the signal.~We employ a state-of-the-art eccentric waveform model incorporating higher-order modes \citep{SEOBNRv5EHM}, and construct an amortized and interpretable WST-based classifier for eccentricity detection.

The WST is a signal-processing technique that builds a multi-scale, multi-orientation representation of the data in the form of coefficients that capture key structural features \cite{mallat2012, bruna2012}. The WST has already been demonstrated as a powerful tool to classify and characterise glitches in interferometer strain data, leading to improvements in robust signal detection \citep{2025PhRvD.111h4044L}. Outside of GW data analysis, the WST has been used in cosmology, see \citep{cheng2021, Cheng__2020,Valogiannis_2022, Valogiannis_2024, Shimabukuro_2025}, as well as metrology, finance and image analysis, see e.g.~\citep{morel2023a, morel2023b,Cheng_2024,angles2018, siahkoohi2023}.

% In this work, we find that the WST coefficients show promise for extracting the characteristic features of GW signals from eccentric CBCs in realistic noise. We train and test three different classifier architectures: logistic regression, one-dimensional CNN, and two-dimensional CNN. 
In this work, we find that the WST coefficients show promise for extracting the characteristic features of GW signals from eccentric CBCs in realistic noise. We develop a one-dimensional CNN (1D-CNN) and compare it against classifiers of varying complexity, namely a logistic regression (LR) and a 2D-CNN. While these methods will likely benefit from future improvements\textemdash such as altering our reference frequency-based definition of eccentricity to one that defines eccentricity at a fixed number of cycles before merger\textemdash we find that they are already able to correctly distinguish between eccentric and quasi-circular signals in the majority of the test set, even when the quasi-circular signals contain spin-induced precession.  

This paper is organized as follows. In Sec.~\ref{sec:methods}, we introduce the dataset used in this study and the wavelet scattering transform formulation. We also present our numerical set-up; a 1D-CNN classifier for the WST coefficients. In Sec.~\ref{sec:ecc_classi}, we present our numerical experiments and results. We conclude in Sec.~\ref{sec:discussion} by discussing the implications of our findings and outlining future directions.

\section{Methods}
\label{sec:methods}

\subsection{Waveform Generation: \texttt{SEOBNRv5EHM}}
\label{sec:training data}

We generate synthetic strain data comprising simulated GW waveforms injected into detector noise, using the \texttt{Bilby} library~\citep{bilby1, bilby2}. Eccentric waveforms are produced using the \texttt{SEOBNRv5EHM} approximant~\cite{gamboa2025}, an effective-one-body, multipolar waveform model describing eccentric binary black hole (BBH) systems with component spins aligned with or against the orbital angular momentum. Priors on the intrinsic parameters are defined within \texttt{Bilby}. The dimensionless component spin magnitudes are drawn from aligned-spin priors spanning nearly the full physical range, $a \in [0,0.99]$, and the relativistic anomaly is sampled uniformly in $[0,2\pi]$. Component masses $m_{1,2}$ are uniformly sampled in $[10,40]\,M_\odot$, while the luminosity distance is sampled through the default \texttt{Bilby} prior in the range $ [100,800]\,\mathrm{Mpc}$\textemdash ensuring a distribution that is uniform in comoving volume and source frame time. We initialise the detectors Gaussian noise coloured by O4 noise curves ("realistic noise" hereafter).

The waveform model-defined eccentricity at a reference frequency of 10~Hz, $e_{10}$, is sampled uniformly within each class: the classification task is to distinguish eccentric systems ($e_{10} \in [0.01,0.5]$) from non-eccentric systems ($e_{10} \in [0.001,0.01]$). While a log-uniform prior is commonly adopted in GW data analysis, uniform sampling ensures balanced class populations during training. Alternative definitions of eccentricity that do not rely on a reference frequency are also possible, e.g.,\citep[]{Vijaykumar_2024, Shaikh_2023}.

Waveforms are generated in the frequency domain over the range $10\,\mathrm{Hz}$--$1024\,\mathrm{Hz}$, where the upper cutoff corresponds to half the sampling rate $f_s$. Each signal has a duration of $128\,\mathrm{s}$ and is transformed to the time domain via an inverse Fast Fourier Transform (FFT). The geocentric merger time is sampled uniformly from $126 \pm 0.1\,\mathrm{s}$, and all waveforms are aligned at merger time. Unless otherwise stated, all six multipoles \footnote{These correspond to $(\ell,m)=(2,2), (2,1), (3,2), (3,3), (4,3)$, and $(4,4)$.} available in \texttt{SEOBNRv5EHM} are included.

The simulated signals are injected into noise realizations of the two Advanced LIGO detectors and the Advanced Virgo detector. An 8\,s window around the estimated peak at merger, $t_m$, is extracted, defined as $[t_m - 7.8\,\mathrm{s},\, t_m + 0.2\,\mathrm{s}]$.~The resulting strain data segments are whitened, and the network signal-to-noise ratio (NSNR) is computed across the detector network. Samples with NSNR $<15$ are discarded to ensure confident detectability while preserving realistic detector noise conditions for training and evaluation. 

For this study, we generate approximately $1.2\times10^5$ waveforms and compute their compressed WST representations. The resulting dataset consists, for each signal and detector, of the scattering coefficients $C$ and the associated time bins $T_b$, where $T_b \ll T \cdot f_s$ and $T=8$s is the total duration of the signal segment considered.

\subsection{Wavelet Scattering Transform}
\label{sec:WST}
 The wavelet scattering transform (WST) is a multi-resolution signal analysis algorithm~\cite{bruna2012,mallat2012,Anden_2014,morel2023a}: a hierarchical filtering framework that combines wavelet convolutions with modulus and averaging operations to build a multiscale representation of a signal. The resulting scattering coefficients are stable to small deformations, including time warping, and preserve higher-order statistical structure, enabling an informative yet compact characterization of the data. The coefficients thus encode localised structure and cross-scale dependencies, providing sensitivity to morphological waveform features such as those induced by orbital eccentricity.

In what follows, we present the key aspects of the one-dimensional WST formulation \cite{Anden_2014} employed in this study. In practice, we implement the transform using the \texttt{Kymatio} package~\cite{andreux2022}, which efficiently computes the required convolutions in the Fourier domain via the FFT.

In one dimension, the WST is parameterised by two quantities, $J$ and $Q$. Given a sampling frequency $f_s$, the parameter $J$ sets the maximum (averaging) scale
\begin{equation}\label{eq: averaging scale}
\tau_J = \frac{2^J}{f_s},
\end{equation}
that is, the largest time scale over which the scattering coefficients are averaged. Increasing $J$ therefore increases translation invariance while reducing temporal resolution. The parameter $Q$, on the other hand, specifies the number of wavelets filters per octave and thus determines the sampling density along the frequency axis---larger values of $Q$ provide a finer discretisation of scales and hence higher frequency resolution.

The first-order scattering coefficients, $S_1$, are obtained by convolving the signal with a wavelet $\psi_{j,q}$, applying the modulus, and a low-pass, $\phi_J$, averaging:
\begin{equation}
S_1(t,j_1,q_1)=
\left( \left| h \ast \psi_{j_1,q_1}  \right| \ast \phi_J \right)(t). \label{eq:s1}
\end{equation}
These coefficients measure the averaged amplitude of signal fluctuations localized near the scale-frequency index $(j_1,q_1)$ while suppressing high-frequency temporal variations.

Second-order scattering coefficients, $S_2$, are computed recursively by applying a second wavelet transform:
\begin{equation}
S_2(t,j_1,q_1,j_2,q_2)=\left(\left|
\left| h \ast \psi_{j_1,q_1} \right| \ast \psi_{j_2,q_2}
\right|
\ast \phi_J
\right)(t), \label{eq:s2}
\end{equation}
with the constraint $j_2 > j_1$, to ensure that the second order wavelet probes amplitude modulations occurring at longer timescales than those captured by the first wavelet. Hence, $S_2$ coefficients recover information lost during the first averaging step and encode correlations between structures at different scales. Averaging with $\phi_J$ ensures that all scattering coefficients are invariant to translations up to a scale $\tau_J$. We refer the reader to ref.~\cite{Anden_2014} and App.~\ref{Ap:1} for a more detailed explanation on the one-dimensional wavelet scattering transform.

\begin{figure*}
    \centering
    \includegraphics[width=0.9\linewidth]{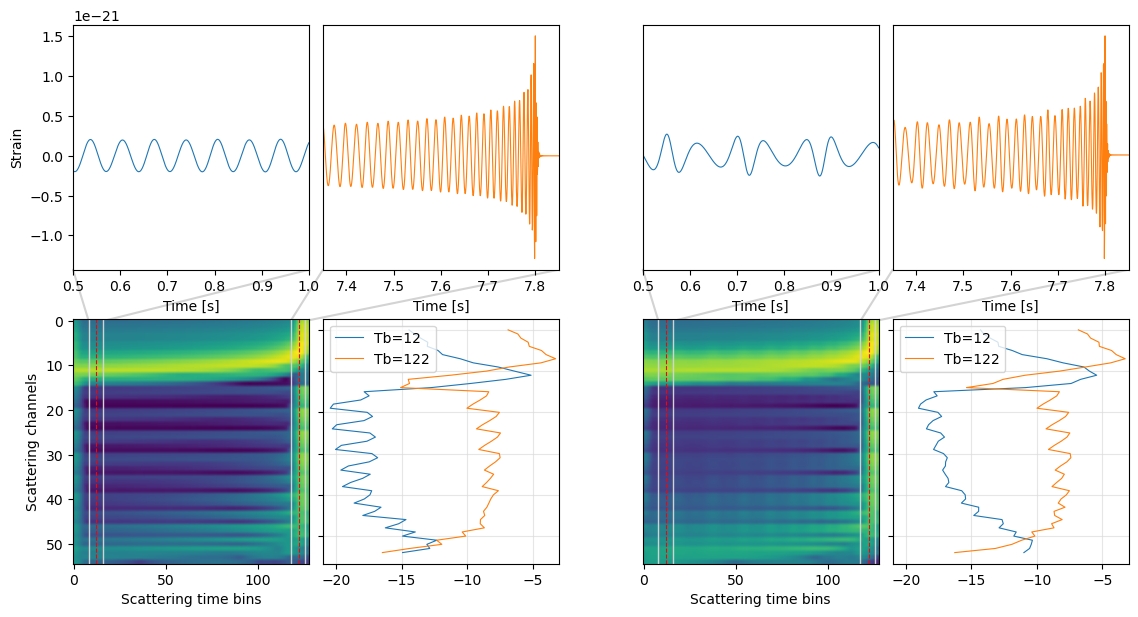}
    \caption{Illustrative example of a \textit{(left 4 panels)} quasi-circular and \textit{(right 4 panels)} eccentric ($e_{10}=0.3$) waveform, with zero noise, and their corresponding $(J,Q) = (7,2)$ WST coefficients (on a logarithmic scale). The binary masses are 12 and 18 $M_\odot$, at a luminosity distance of $100$ Mpc. The top row shows two $0.5$~s windows of the waveform, where the first represents the inspiral and the second the merger. The bottom row represents all WST coefficients in the left panel, as well as a slice at two time bins $T_b$ in the right panel, corresponding to the different phases in the top row. From these plots, it is clear that, as expected, the WST coefficients representing the eccentric binary distribute the power across the channels more evenly; i.e.,~many of the WST coefficients have increased values compared to the circular case. It should be noted that the top 15 channels in the bottom row represent first-order scattering coefficients, and the rest second-order.}% Additionally, the above picture can be heavily obscured in the presence of noise --- see Fig.~\ref{fig: noisy data} in the Appendix.}
    \label{fig: clean data}
\end{figure*}

In a similar way, higher-order coefficients can be defined recursively. However, their amplitudes decrease rapidly with order, and most of the relevant information is captured by the first two orders~\cite{mallat2012}. Accordingly, in this work we restrict our analysis to orders $\le 2$. Figure \ref{fig: clean data} shows that the WST representation captures higher-order structure and amplitude modulations, including those induced by orbital eccentricity.

\subsection{Numerical Set-Up: 1D-CNN}
\label{sec:classifiers}

CBCs with non-zero orbital eccentricity produce GW signals whose time-frequency structure differs from that of quasi-circular inspirals~\cite{patterson2024,islam2024b,Vijaykumar_2024}. Eccentric systems radiate power into multiple harmonics of the orbital frequency, producing a richer multi-resolution structure that the WST is able to capture \cite{bruna2011}. Figure~\ref{fig: clean data} illustrates this by comparing the WST coefficients of a circular and an eccentric system.

In what follows, we present a one-dimensional convolutional neural network (1D-CNN) classifier designed to analyse the combined WST coefficients from the individual GW detectors and to exploit their multi-resolution structure for eccentricity classification. Our main goal is to develop a parsimonious strategy that reduces computational complexity and thereby speeds up classification: by compressing the signals and working with lower-dimensional classifiers, we avoid the need for more costly intermediate steps such as constructing spectrograms and training complex CNN architectures. We benchmark our approach againts two other classifiers of contrasting complexity: a linear model (Logistic Regression, LR) and a non-linear model (2D-CNN), an architecture originally designed for image analysis, which we discuss in Appendix~\ref{app: other classifiers}.

\subsubsection{One-dimensional CNN}
\label{sec:1dcnn}

\begin{figure*}[t] 
\centering
\begin{tikzpicture}[
  scale=0.75, transform shape,   
  font=\scriptsize,              
  arrow/.style={-Latex, thick},
  block/.style={
    draw, rounded corners=2pt,
    align=center,
    minimum height=10mm,          
    minimum width=20mm,         
    inner sep=2pt
  },
  lane/.style={
    draw, dashed, rounded corners=3pt,
    inner sep=4pt
  },
  note/.style={font=\tiny, align=center},
  node distance=6mm and 8mm     
]
% -------------------------
\node[block] (inp){Input\\ $(B,D,C,T)$};
\node[note, right=5mm of inp] (loop){};
\node[block, right=5mm of loop] (slice){Slice\\ $(B,C,T)$};
\node[block, right=6mm of slice, minimum width=26mm] (cnn){CNN Encoder\\ $(B,E,L)$};
\node[block, right=6mm of cnn] (pool){AvgPool$(T_p)$\\ $(B,E,T_p)$};
\node[block, right=6mm of pool] (flat){Flatten\\ $(B,F)$};
\node[lane, fit=(loop)(flat),label={[note]above:Per-detector}] (lane1) {};
% -------------------------
\draw[arrow] (inp) -- (slice);
\draw[arrow] (slice) -- (cnn);
\draw[arrow] (cnn) -- (pool);
\draw[arrow] (pool) -- (flat);
% -------------------------
\node[block, right=6mm of flat] (stack){Stack\\ $(B,D,F)$};
\node[block, below=10mm of stack, minimum width=24mm] (stats){StatsPool\\ $(B,3F)$};
\node[block, right=7mm of stats, minimum width=24mm] (mlp){MLP\\ $(B,1)$};
\node[block, right=7mm of mlp] (out){Logit};
\node[lane, fit=(stack)(stats),label={[note]above:Fusion}] (lane2) {};
\node[lane, fit=(mlp)(out),label={[note]above:Head}] (lane3) {};
% -------------------------
\draw[arrow] (flat) -- (stack);
\draw[arrow] (stack) -- (stats);
\draw[arrow] (stats) -- (mlp);
\draw[arrow] (mlp) -- (out);
\end{tikzpicture}
\caption{Block diagram of the 1D-CNN designed for eccentricity detection. Data from different detectors are encoded separately by a shared one-dimensional convolutional layer with three blocks and intermediate max-pooling for temporal downsampling. Feature maps are averaged into $T_p$ bins, producing fixed-length detector embeddings. These are fused (joined) using first order statistics, and the resulting representation is classified by a shallow MLP to yield the final probability value.}
\label{1dcnn}
\end{figure*}

A key consideration is that each detector in the network has distinct characteristics, most notably different noise power spectral densities and antenna response patterns. To account for these differences while still learning common signal features, the architecture, Fig.~\ref{1dcnn}, takes the WST representation of the data from each detector as input and applies a shared convolutional block to their WST features, ensuring that the learned filters capture detector-independent temporal signatures. The resulting detector embeddings are subsequently fused into a joint representation that captures the common source information encoded in their data. Concretely, the input to the 1D-CNN is a WST tensor of shape
\(
\bm{X}_n\in\mathbb{R}^{B\times D\times C\times T_b},
\)
where \(B\) is the data batch size, \(D\) the number of detectors, \(C\) the number of WST frequency channels, and \(T_b\) the number of time bins. For each detector \(d\), the slice \(\bm{X}_{n,d}\in\mathbb{R}^{B\times C\times T_b}\) is processed independently.\\
 
To make the time axis consistent across events on the same detector, the synthetic data are aligned at merger. This alignment ensures that corresponding temporal features occur at the same relative positions in the scattering representation, allowing the network to learn consistent temporal patterns associated with eccentricity. Although times of arrival are slightly offset between detectors for the same event, as is also the case for real detections, this effect is mitigated by the modulus operation of Eqs.~\eqref{eq:s1}-\eqref{eq:s2}. 

The network maps the WST data into a lower-dimensional representation for each event and detector across channels and time bins. It consists of three convolutional blocks, each composed of a 1D-convolutional layer followed by batch normalisation and a \emph{ReLU} activation, with \emph{max-pooling} applied after the first two blocks to reduce the temporal dimension. This encoder structure progressively reduces the number of feature channels: $C \rightarrow 64 \rightarrow 32 \rightarrow 16$, using kernel sizes $(7,5,3)$. The resulting encoding is
\begin{equation}
\bm{H}_{n,d} \doteq f_{\mathrm{cnn}}(\bm{X}_{n,d}) \in \mathbb{R}^{E \times L},
\end{equation}
where $E=16$ is the number of embedding channels and $L=\mathrm{floor}(T_b/2)$ denotes the reduced temporal dimension. To obtain a fixed-length, merger-aligned summary, we apply adaptive average pooling to $T_p$ time bins,
\begin{equation}
\tilde{\bm{H}}_{n,d} = \mathrm{Pool}_{T_p}(\bm{H}_{n,d}) \in \mathbb{R}^{E \times T_p}\;,
\end{equation}
effectively dividing the time axis into $T_p$ coarse temporal segments and producing a compact representation of the scattering coefficients. The pooled representation is then flattened to yield the final detector embedding
\begin{equation}
\bm{e}_{n,d} = \mathrm{vec}\!\left(\tilde{\bm{H}}_{n,d}\right) \in \mathbb{R}^{E\cdot T_p}.
\end{equation}
This feature representation captures learned temporal correlations across WST channels at multiple time scales and across detectors.

To construct a joint detector representation, we stack the individual detector embeddings into a detector-set representation
\begin{equation}
\bm{Z}_n = [\bm{e}_{n,1},\ldots,\bm{e}_{n,D}] \in \mathbb{R}^{D \times (E\cdot T_p)},
\end{equation}
and pool across the detector dimension to form the final feature vector \(\bm{p}_n \in \mathbb{R}^{3(E\cdot T_p)}\). This vector is then passed to a two-layer multilayer perceptron (with ReLU activation and dropout of 0.3), which outputs a scalar, s, defining the eccentric versus non-eccentric classification as explained in the Sec.~\ref{sec: CV-TC-E}. With default parameters---$T_p = 6$, an intermediate multilayer perceptron (MLP) of size 128, and $(J,Q) = (7,2)$---the model has 74\,321 trainable parameters.

\subsubsection{Cross-Validation, Threshold Calibration, and Evaluation}
\label{sec: CV-TC-E}
To evaluate the performance of the 1D-CNN approach, we reserve a stratified hold-out test set comprising 20\% of the data, which is excluded from both training and threshold selection. This ensures that neither the model parameters nor the operating thresholds are influenced by the test data. The remaining 80\% is used for stratified $K$-fold cross-validation with $K=10$, reducing sensitivity to the choice of validation split.

The network is trained as binary classifier, eccentric ($e_{10} \in [0.01,0.5]$) versus non-eccentric ($e_{10} \in [0.001,0.01]$), by minimising the binary cross-entropy loss. The threshold value $e_{10}^{\mathrm{thr}} = 0.01$ adopted for ``circular'' events is a practical choice. Optimisation is performed using the Adam optimiser with a learning rate of $5 \times 10^{-4}$ and a batch size of 1000. To mitigate overfitting and promote generalisation, early stopping is applied independently within each fold based on the validation AUC (area under the receiver operating characteristic curve), with a patience of five epochs.

Since signals confidently identified as eccentric are expected to constitute only a small fraction of detected CBCs, we evaluate the model at fixed operating points along the false-alarm-rate (FAR) axis---also known in the machine-learning literature as the false positive rate (FPR). Fixing the FPR enables a fair comparison across classifiers, since the same score threshold can correspond to very different classification rates for different classifiers. We define the FPR as the fraction of negative samples assigned to the positive class,
\begin{equation}
    \text{FPR} = \frac{\text{FP}}{\text{FP} + \text{TN}} = \Pr(\hat{y}=1 \mid y=0).\label{eq:far}
\end{equation}
Note that we use FPR in the dimensionless classification sense (the fraction of negatives misclassified as positives), not the per-unit-time event rate conventional in LVK searches. At a fixed FPR, we quantify performance by the fraction of positive samples correctly recovered, the true positive rate (TPR),
\begin{equation}
    \text{TPR} = \frac{\text{TP}}{\text{TP} + \text{FN}}
    = \Pr(\hat{y}=1 \mid y=1).\label{eq:tpr}
\end{equation}
For each fold, the decision threshold $s^\star$ is calibrated on the validation set against the distribution of negative scores: to achieve a target FPR of $\alpha$, we set the threshold $s^\star$ to the $(1-\alpha)$ quantile of the negative validation scores. For example, for a target FPR of $1\%$, $s^\star$ is chosen so that $99\%$ of the negative validation scores fall below it. The calibrated thresholds are then applied to the held-out test set to measure the realised FPR and TPR; any deviation of the realised FPR from its target reflects how well the threshold generalises from validation to test data. Finally, to quantify overall discriminative performance independent of any specific operating point, we additionally report the AUC and the average precision (AP). In App.~\ref{app: other classifiers} we report the same metrics for the LR and 2D-CNN classifiers for comparison.

\section{Numerical experiments and results}
\label{sec:ecc_classi}
In this section, we present numerical experiments evaluating eccentricity detection performance using the WST of the simulated GW strain data described in Sec.~\ref{sec:training data}. The primary results are obtained with the 1D-CNN (Sec.~\ref{sec:1dcnn}). For completeness and comparison, we also report results obtained with a 2D-CNN classifier and LR in App.~\ref{app: other classifiers}.
\begin{table}[t]
\centering
\caption{Comparison of detection performance for WST data computed on the same dataset, but with $Q=1, 2$ and fixed $J=7$. Shown are the mean and standard deviation across $K=10$ folds for $\text{FPR} =\{0.1, 0.01, 0.001\}$. Results are obtained with the 1D-CNN model. The relative difference is given by $\Delta = \big((M_{Q=2}-M_{Q=1})/M_{Q=1}\big)\times 100$ computed using the column means.}
\label{tab:swt_q_comparison_j7}
\begin{tabular}{lccc}
\hline
 & \textbf{Q = 2} & \textbf{Q = 1} & $\boldsymbol{\Delta\%}$ \\
\hline
Validation AUC &
$0.829 \pm 0.010$ &
$0.823 \pm 0.004$ &
+0.73 \\
Validation AP &
$0.862 \pm 0.010$ &
$0.856 \pm 0.004$ &
+0.70 \\
Test AUC &
$0.831 \pm 0.011$ &
$0.824 \pm 0.003$ &
+0.85 \\
Test AP &
$0.863 \pm 0.010$ &
$0.856 \pm 0.003$ &
+0.82 \\
\hline
\hline
Validation TPR\\
\hline
FPR 0.1 &
$0.599 \pm 0.023$ &
$0.580 \pm 0.011$ &
+3.28 \\
FPR 0.01 &
$0.422 \pm 0.026$ &
$0.408 \pm 0.014$ &
+3.43 \\
FPR 0.001 &
$0.333 \pm 0.033$ &
$0.326 \pm 0.025$ &
+2.15 \\
\hline
\hline
Test TPR\\
\hline
FPR 0.1 &
$0.600 \pm 0.028$ &
$0.585 \pm 0.010$ &
+2.56 \\
FPR 0.01 &
$0.424 \pm 0.026$ &
$0.411 \pm 0.012$ &
+3.16 \\
FPR 0.001 &
$0.337 \pm 0.032$ &
$0.331 \pm 0.024$ &
+1.81 \\
\hline
\end{tabular}
\end{table}
\begin{table}[t]
\centering
\caption{Performance of the 1D-CNN classifier for different choices of the WST parameters $J$ and $Q$ in the data. Performance is reflected by the threshold-agnostic metrics AUC and AP on the test set, using the best fold. Finally, the TPR is given for a target FPR of 10\%.
}
\label{tab: J Q comparison 1D CNN}
\begin{ruledtabular}
\begin{tabular}{lccc}
$\mathbf{(J,Q)}$ & \textbf{AUC} & \textbf{AP} & \textbf{TPR$_{0.1}$} \\
\hline
$\mathbf{(7,2)}$
%& $\sim50^{\ast}$s
& $\mathbf{0.844}$ & $\mathbf{0.876}$ & $\mathbf{0.636}$ \\
$(8,2)$
%& $\sim1^{\ast}$m
& $0.832$ & $0.863$& $0.597$ \\
$(6,2)$
%& $\sim1.5^{\ast}$m
& $0.845$ & $0.876$& $0.622$
\end{tabular}
\end{ruledtabular}
\end{table}
Across our experiments, the 1D-CNN architecture applied to WST coefficients with $(J,Q)=(7,2)$ emerges as the best-performing configuration overall, and we adopt it as our reference model in what follows. We note that a suitable choice of $J$ depends strongly on the length of the data considered and the sampling rate; see App.~\ref{Ap:1} for a detailed explanation.
Metrics for this configuration are reported in Table~\ref{tab:swt_q_comparison_j7}, 
for $\text{FPR} ={10^{-1}, 10^{-2}, 10^{-3}}$---alongside results for $(J,Q)=(7,1)$, which illustrate the benefit of increasing the scattering depth $Q$ to incorporate additional frequency-dependent structure. Further decreasing $J$ or increasing $Q$ can yield modest improvements, but at the cost of a rapidly growing WST representation (see Table \ref{tab: J Q comparison 1D CNN} reporting best-fold metrics for each model); the choice $(J,Q)=(7,2)$ shows the best trade-off between performance and feature dimensionality. Further validation metrics are shown in App.~\ref{app: other classifiers}.% where a confusion matrix for a target FPR of 10\%, as well as the ROC curve in Figs.~\ref{fig:cm_j7q2} and \ref{fig: 1dcnn_ref_test_roc}.
\subsection{Characterising Eccentric Systems}
\label{sec: results, distr of FN}
\begin{figure*}[ht!]
    \centering
    \includegraphics[width=0.95\linewidth]{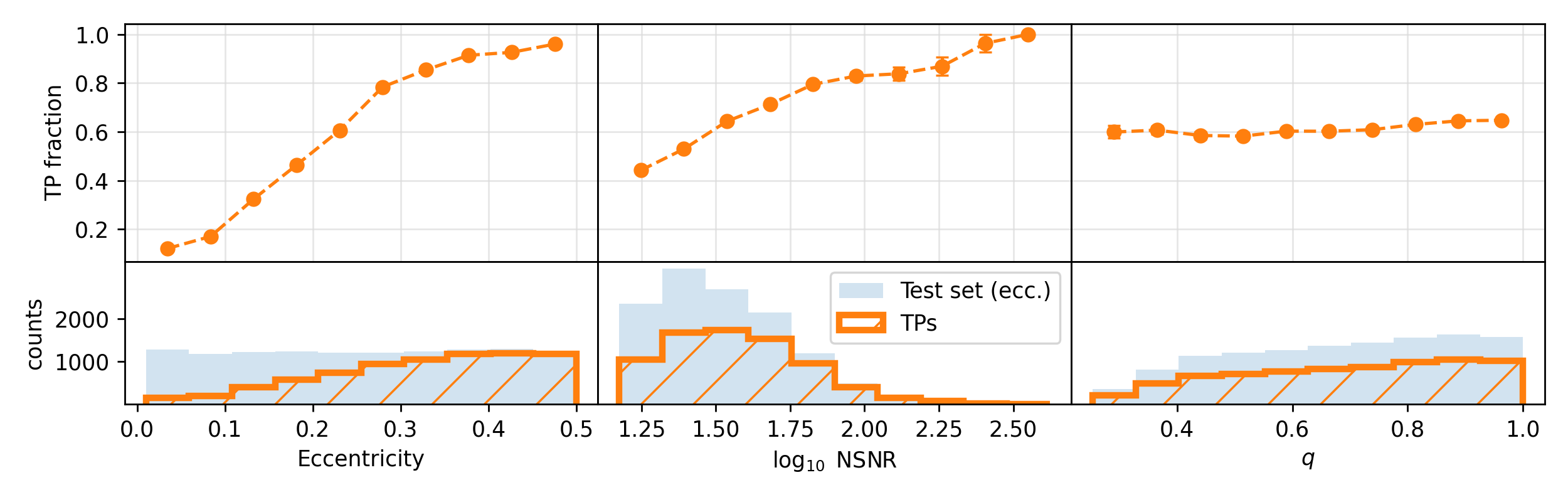}
    \caption{Performance of the reference 1D-CNN classifier, with the decision threshold $s^\star$ calibrated to a FPR of $10\%$ (Sec.~\ref{sec: CV-TC-E}), on the eccentric binaries in the test set. The performance is shown for different bins of the \textit{(left)} eccentricity $e$, \textit{(middle)} network SNR and \textit{(right)} mass ratio $q$. The top row shows the fraction of true positives (TPs) in each bin, the bottom row shows a histogram of the test set and the TPs.}
    \label{fig: performance e snr q}
\end{figure*}
Weakly eccentric systems---i.e.~binaries with eccentricities close to the class threshold $e_{10}^\mathrm{thr} = 0.01$---are expected to be the most difficult to distinguish from quasi-circular systems, and therefore the most likely to be misclassified. To quantify this, we focus on the eccentric binaries in the test set ($e_{10} > e_{10}^\mathrm{thr}$) and extract the subset of true positives (TP), i.e.~the eccentric binaries correctly classified as eccentric at a target FPR of $10\%$. These TPs are the eccentric binaries whose classification score $s > s^\star$ calibrated to this FPR (Sec.~\ref{sec: CV-TC-E}). The left panel of Fig.~\ref{fig: performance e snr q} shows that the eccentricity distribution of the TP subset is skewed toward higher values of $e_{10}$, consistent with the expectation that weakly eccentric systems are harder to distinguish from circular ones. The classifier thus identifies highly eccentric binaries with greater confidence, with misclassifications concentrated near the class boundary.

\begin{figure}[t]
  \centering
  \includegraphics[width=\columnwidth]{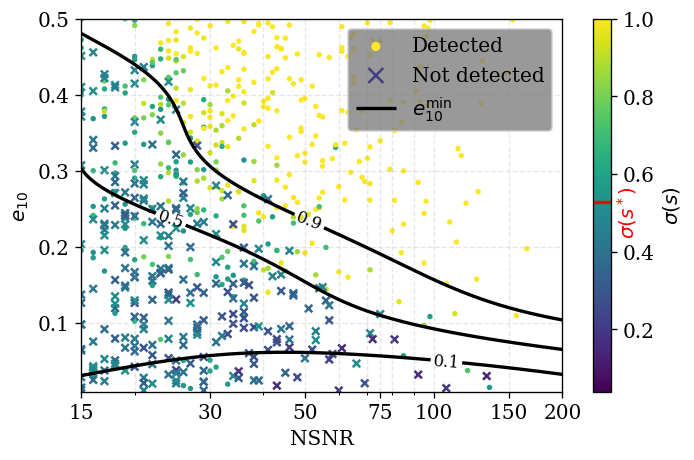}
  \caption{Estimate of the minimum detectable eccentricity $e^{\min}_{10}$,
  Eq.~\eqref{eq:emin}, at detection probabilities $P_{\rm det}=0.9, 0.5, 0.1$ as a function of NSNR, for a fixed $\mathrm{FPR}=10\%$. The colorbar corresponds to the classifier output $s$, rescaled to $[0,1]$ with a sigmoid $\sigma$, and $s^\star$ is the decision threshold for this FPR (Sec.~\ref{sec: CV-TC-E}). The points correspond to a sample of the data: dots (crosses) indicate that the binary was (in)correctly classified as eccentric (circular) by the reference 1D-CNN.}
  \label{fig:emin_snr}
\end{figure}

The strength of a GW signal depends primarily on the binary masses and luminosity distance, which together set the SNR and, in turn, the classifier performance. To investigate how these physical parameters influence eccentricity classification, we examine the dependence on the total mass $M$, the mass ratio $q = m_2/m_1 \leq 1$, and the network SNR (NSNR). Figure~\ref{fig: performance e snr q} shows the distribution of TPs as a function of NSNR and $q$: the mass ratio has only a minor effect on performance, whereas the NSNR is highly influential, with low-SNR events exhibiting substantially lower TP rates. The influence of the total mass is discussed further in Sec.~\ref{sec: results, def of ecc}.
In addition, we characterise classifier performance through the minimum detectable eccentricity $e^{\min}_{10}$. For a given NSNR, we define it as the smallest eccentricity at which the TPR reaches a target detection probability $P_{\rm det}$:
\begin{equation}
  e^{\min}_{10}({\rm NSNR}) := \inf \big\{e \,  \big| \, {\rm TPR(e, NSNR) \geq P_{\rm det}}\big\}\,.
  \label{eq:emin}
\end{equation}
Estimating $e^{\min}_{10}$ requires a smooth model of the detection probability across the $(e,\,\mathrm{NSNR})$ plane. To this end, we apply the \texttt{GaussianProcessClassifier} (GPC) from \texttt{scikit-learn}~\cite{scikit-learn}: as input, we randomly sample 2500 eccentric binaries from the test set\footnote{We do not use the full test set due to computational overhead. We checked that increasing the number of samples does not significantly alter the results.} and use the corresponding values of $e$ and $\log_{10}({\rm NSNR})$.
The labels encode the detection outcome: a value of 1 indicates that an eccentric binary was correctly identified by the 1D-CNN classifier at the fixed FPR (i.e.,~its score $s$ exceeds $s^\star$; see Sec.~\ref{sec: CV-TC-E}), and 0 otherwise. The GPC fits a continuous latent function $f\!\left(e, \log_{10}({\rm NSNR})\right)$ and maps it to a detection probability through a logistic sigmoid. We adopt a composite kernel---a constant kernel multiplied by a radial basis function (RBF) kernel with an independent length-scale for each input---and optimise its hyperparameters by maximising the marginal log-likelihood during fitting. The minimum detectable eccentricity at a fixed detection probability $P_{\rm det}$ is then given by the corresponding isocontour of the fitted detection-probability surface. As shown in Fig.~\ref{fig:emin_snr}, $e^{\min}_{10}$ decreases monotonically with NSNR, as expected, for both $P_{\rm det} = 0.9$ and $P_{\rm det} = 0.5$.

\subsection{Contribution of Higher Multipoles}

To investigate the impact of higher multipoles on eccentricity detection, we train the 1D-CNN classifier on \texttt{SEOBNRv5EHM} waveforms restricted to the dominant quadrupole mode, $(\ell,m)=(2,2)$, and evaluate its performance on data that include all multipoles. The results are summarised in Table~\ref{tab:ml_comparison_errors_full} for different FPRs, $10^{-1}, 10^{-2}, 10^{-3}$. We can see that training on quadrupole-only waveforms the WST representation yields robust classification performance, with test AUC $\approx 0.83$ and stable detection efficiency (TPR) across all operating points. This indicates that the quadrupole mode captures the dominant eccentricity-dependent structure present in the scattering coefficients. We find that the quadrupole-only and all-modes models exhibit statistically consistent performance across all metrics. The all-modes model achieves marginally lower test AUC and AP, with relative differences of $0.36\%$ and $0.23\%$, respectively, but these are small compared to the fold-to-fold variation and do not indicate a statistically significant improvement\footnote{These findings are consistent with \cite{tang2026} published after our work was submitted.}. The true-positive rate  at fixed FPR are similarly consistent. On the validation set, the quadrupole-only model yields slightly higher TPR at all FPR thresholds, while on the independent test set (which includes all modes) the two models agree to within $\lesssim 1.45\%$ across all operating points.

The ability of the 1D-CNN to achieve strong performance using only quadrupole information reflects its capacity to extract nonlinear temporal features from the scattering coefficients. At the same time, the absence of improvement from including higher multipoles suggests that their contribution to eccentricity detection is subdominant in the mass range $10$--$40\,M_{\odot}$. 

\begin{table}[t]
\centering
\caption{Performance comparison between the quadrupole-only and full (all modes) 1D-CNN model, for $\text{FPR} =\{0.1, 0.01, 0.001\}$. Results are reported over $K=10$ folds, as mean $\pm$ standard deviation. The relative difference is defined as $\Delta = \big[(M_{\mathrm{quad}}-M_{\mathrm{all}})/M_{\mathrm{all}}\big]\times 100$, computed from the column means.}
\label{tab:ml_comparison_errors_full}
\begin{ruledtabular}
\begin{tabular}{lccc}
 & Quadrupole & All modes & $\Delta(\%)$ \\
\hline
Validation AP    & $0.864 \pm 0.014$ & $0.862 \pm 0.010$ & $+0.23$ \\
Test AUC         & $0.828 \pm 0.013$ & $0.831 \pm 0.011$ & $-0.36$ \\
Test AP          & $0.861 \pm 0.012$ & $0.863 \pm 0.010$ & $-0.23$ \\
\hline
Validation TPR \\
\hline
FPR 0.1          & $0.61 \pm 0.03$ & $0.60 \pm 0.02$ & $+1.67$ \\
FPR 0.01         & $0.43 \pm 0.04$ & $0.42 \pm 0.03$ & $+2.38$ \\
FPR 0.001        & $0.34 \pm 0.04$ & $0.33 \pm 0.03$ & $+3.03$ \\
\hline
Test TPR \\
\hline
FPR 0.1          & $0.60 \pm 0.03$ & $0.60 \pm 0.03$ & $+0.00$ \\
FPR 0.01         & $0.43 \pm 0.04$ & $0.42 \pm 0.03$ & $+2.38$ \\
FPR 0.001        & $0.34 \pm 0.04$ & $0.34 \pm 0.03$ & $+0.00$ \\
\end{tabular}
\end{ruledtabular}
\end{table}

\subsection{Binaries with Spin-Induced Precession}

\begin{figure}[t]
    \centering
    \includegraphics[width=0.9\linewidth]{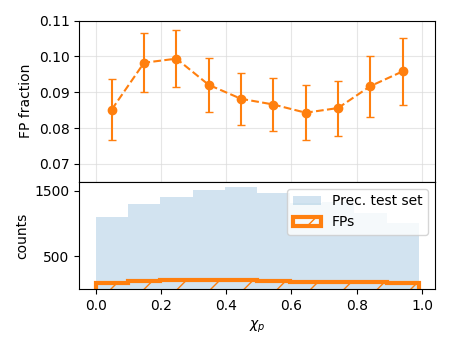}
    \caption{\textit{(bottom)} Histogram of $\chi_p$ values in the precessing, quasi-circular binary dataset, together with the subset classified as false positives (FPs) by the reference 1D-CNN tuned to a FPR of $10\%$. \textit{(top)} Fraction of FPs in each bin of the bottom panel.}
    \label{fig: FPs in prec data}
\end{figure}

Eccentricity and spin-induced precession can produce qualitatively similar amplitude and phase modulations in GW signals~\citep[e.g.,][]{Romero_Shaw_2023}, so a classifier trained to identify eccentricity may instead respond to precession-induced features. To assess this potential degeneracy, we apply our classifier to a test set of quasi-circular, precessing BBHs generated with the \texttt{SEOBNRv5PHM} approximant~\citep{ramosbuades2023seobnrv5phm}. Both the \texttt{SEOBNRv5PHM} (non-eccentric, precessing-spin) and \texttt{SEOBNRv5EHM} (eccentric, aligned-spin) models share the same aligned-spin, quasi-circular \texttt{SEOBNRv5HM} baseline used during training. This common baseline limits modelling systematics differences between training and test data; in particular, the mismatch between \texttt{SEOBNRv5PHM} and \texttt{SEOBNRv5HM} is\footnote{LVK private communication.} $\sim 10^{-6}$ . We adopt the standard BBH priors defined in \texttt{Bilby}, with the reference frequency for spin definitions fixed at $10$~Hz.

We quantify the degree of spin precession using the effective precession parameter $\chi_p$~\cite{ramosbuades2023seobnrv5phm},
\begin{equation}
    \chi_p = \frac{1}{B_1 m_1^2} \max\left( B_1 m_1^2 \chi_{1,\perp}, \, B_2 m_2^2 \chi_{2,\perp} \right)\,,\label{eq:efchi}
\end{equation}
where $m_1 \geq m_2$ are the component masses, $B_i = 2 + 3 m_{\neq i}/m_i$, and $\chi_{i,\perp}$ is the component of the dimensionless spin perpendicular to the orbital angular momentum. To ensure broad coverage of the precession parameter space, we draw spin magnitudes from $\beta$-distributions with parameters $\alpha=\beta=0.5$, which preferentially sample both small and near-extremal spins and thereby enhance coverage in $\chi_p$.

To understand how precession affects the scattering representation, we examine the first-order scattering coefficients $S_1$,  for precessing waveforms. From Eqs~\eqref{eq:Wmorlet}, \eqref{eq:s1} and \eqref{eq:phm} follows that $S_1$ is bounded by
\begin{equation}
  S_1(\lambda,t)
  \leq C \int dt \int dv \left| \sum_{m'} d^\ell_{m m'}(\beta) \, A_{\ell m'}(t)\, e^{-\frac{(t-v)^2}{2\sigma^2}} \right|,
\end{equation}
where $C$ is a constant. This makes explicit how the mode-mixing coefficients $d^\ell_{m m'}(\beta)$ are captured by the scattering representation. For precessing systems, the precession angle $\beta(t)$ evolves on the precession timescale and generates the additional amplitude and phase modulations characteristic of precessing systems. In the aligned-spin limit relevant to \texttt{SEOBNRv5EHM}, by contrast, $\beta$ is constant and these time-dependent modulations are absent, yielding qualitatively different WST representations in the two cases (see App.~\ref{sec:waveform_structure}).

Figure~\ref{fig: FPs in prec data} shows the fraction of precessing binaries misclassified as eccentric by the reference 1D-CNN classifier, calibrated to a FPR of $10\%$. Across all values of $\chi_p$, the FP fraction remains $\lesssim 10\%$, consistent with the targeted FPR. We note that this fraction does not increase significantly with $\chi_p$, even though precessing systems are absent from the training data. The classifier therefore does not strongly confuse precession-induced modulations with eccentricity signatures\footnote{Similar results were reported by Ref.~\cite{tibrewal2026} after submission of this manuscript}: its false-positive rate stays consistent with the calibrated operating point, indicating that the WST-based representation captures features that distinguish eccentricity from spin-induced precession.\footnote{Repeating the analysis with a target FPR of $1\%$ yields the same conclusion: the FP fraction on the precessing dataset remains consistent with the target FPR.}
\section{Discussion and Conclusions}
\label{sec:discussion}

\begin{figure}[t]
    \centering
    \includegraphics[width=0.9\linewidth]{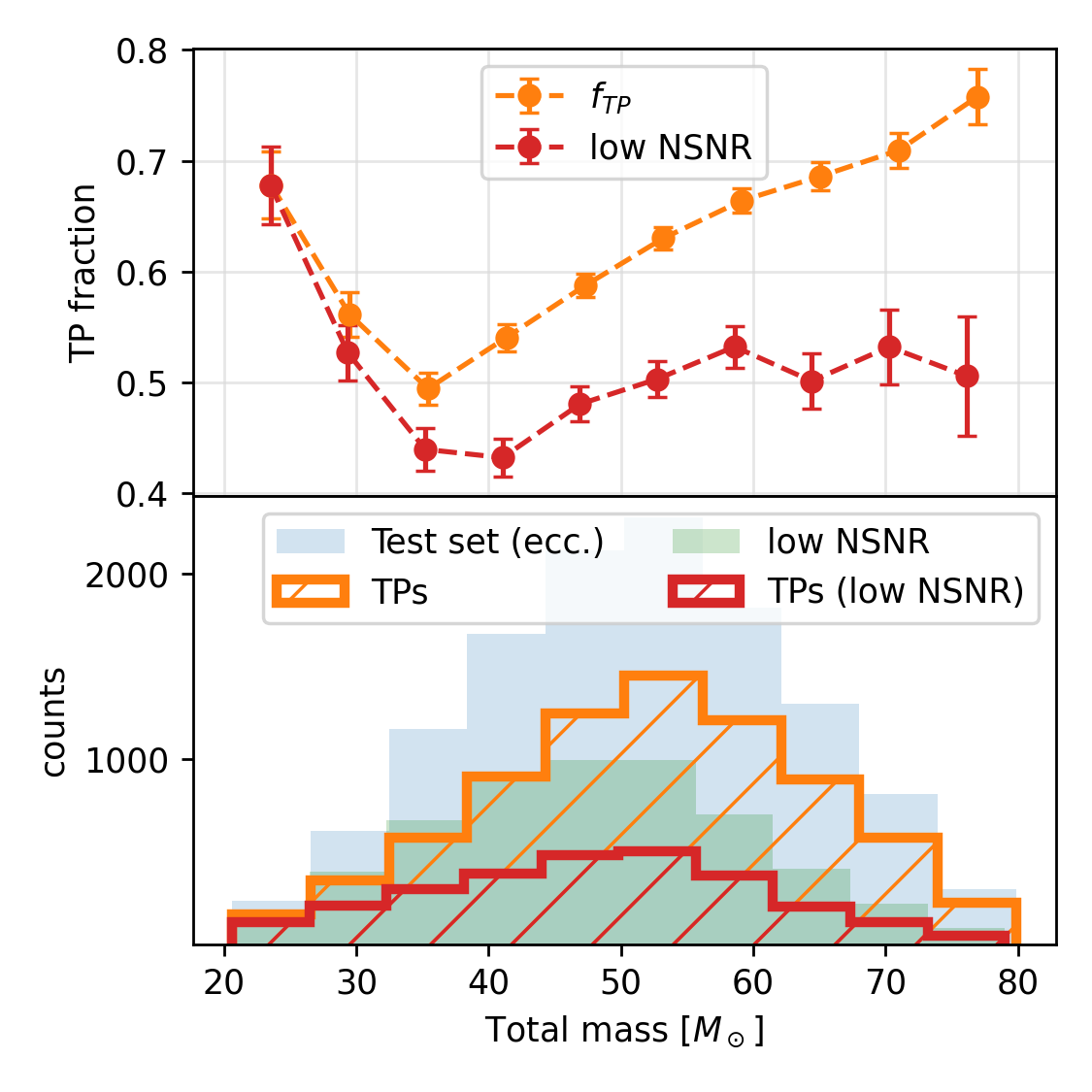}
    \caption{\textit{(bottom)} Distribution of total mass within the test set of eccentric binaries, and within the subset with low NSNR ($<30$). Hatched histograms show the true positives (TPs) in each. \textit{(top)} Fraction of TPs in each bin of the bottom panel, for the full eccentric set and the low-NSNR subset.}
    \label{fig: FN distr TM comp}
\end{figure}

\subsubsection{Definition of Eccentricity}
\label{sec: results, def of ecc}

Eccentricity decreases during the inspiral due to GW emission \citep{P_64}, so its observational signatures are most prominent at earlier times. Lower-mass binaries remain in the detector band for longer before merger, and one might therefore expect eccentricity to be easier to detect in such systems.

Our results initially suggest the opposite: classifier performance appears to improve for higher-mass binaries (Fig.~\ref{fig: FN distr TM comp}). This is explained by a correlation between total mass and NSNR in our dataset, since higher-mass binaries tend to produce stronger signals, in part because of the restricted luminosity-distance range. To disentangle these effects, we examine the TP fraction as a function of total mass while restricting to lower-SNR events (NSNR $<30$), reducing this correlation. Under this restriction, higher-mass binaries become more difficult to classify correctly, as shown in Fig.~\ref{fig: FN distr TM comp}. This is consistent with expectations, since higher-mass binaries have shorter observable inspirals and therefore less visible eccentricity signatures.

This counter-intuitive behaviour as a function of total mass may be exacerbated by the definition of eccentricity used to label the dataset, namely the eccentricity evaluated at a fixed reference frequency of $10$~Hz. We capture only the last $8$~s of the signal in the WST, which, for lower-mass binaries, omits much of the early inspiral. This both reduces the SNR within the analysed portion of the signal and shifts the analysis to a less eccentric part of the waveform, since eccentricity decreases as the orbital frequency increases. As a result, binaries labelled eccentric at $10$~Hz may appear nearly circular over the portion of the signal we pass to the classifier, increasing the likelihood of misclassification. We will mitigate this effect by extending to longer segments in future work, discussed further below. 

\subsubsection{Conclusions and Future Work}
\label{sec:conclu}

In this work, we have demonstrated that the WST provides a compressed, robust, and physically meaningful representation for identifying orbital eccentricity in gravitational-wave signals embedded in realistic detector noise. To exploit these properties, we designed an interpretable and parsimonious 1D-CNN classifier that achieves strong detection performance while remaining computationally efficient, suggesting that the essential eccentricity information is already captured by the wavelet scattering coefficients.

Among the physical factors that influence classifier performance, the signal-to-noise ratio is the dominant driver of the TPR, while the total mass and mass ratio play secondary roles once their correlations with signal strength are controlled for.

We further showed that strong performance can be achieved using quadrupole information alone, indicating that the dominant eccentricity signatures are encoded in the nonlinear features of the scattering coefficients. The lack of improvement from including higher multipoles suggests that their contribution is subdominant in the mass range $10$--$40\,M_{\odot}$, although they are expected to become more important for systems with larger mass ratios or stronger precession-eccentricity correlations.

We have also investigated whether our approach can distinguish eccentricity from spin-induced precession, applying it to a test set of quasi-circular waveforms generated with \texttt{SEOBNRv5PHM}~\cite{SEOBNRv5PHM}, a state-of-the-art model that allows for misaligned spins. We find that classifier performance is consistent across the full range of $\chi_p$, indicating that the WST is an informative representation for separating eccentric from precessing binaries.

This work can be extended in several directions. The proposed framework could be generalised to perform joint detection and parameter estimation and to cover a broader mass range. Alternative eccentricity definitions---better aligned with the observable signal---could also be explored, together with an assessment of performance over wider regions of parameter space. The code to reproduce the WST classification pipeline and example data is publicly available \cite{wst-eccentricity}.

\section{Acknowledgments}
The authors thank Adhrit Ravichandran and Dr Rudy Morel for their helpful comments, and especially Dr Christopher Moore for his valuable insights. We are also grateful to Prof. Carola-Bibiane Sch\"onlieb for her support, and to the Bristol Supercomputing Centre for computational resources. Additional computations were done on the CSD3 cluster (Cambridge). The authors also thank the referees for their thorough and insightful feedback on our manuscript.

P. C. M. is supported by a Daphne Jackson Fellowship sponsored by the Alan Turing Institute. S. J. S. is supported by the Centre for Doctoral Training at the University of Cambridge, funded by STFC.
I. M. R-S acknowledges support from the STFC Ernest Rutherford Fellowship, grant number UKRI2423.

\section{Appendix}
\setcounter{secnumdepth}{2}
\appendix

\section{SEOBNRv5EHM and SEOBNRv5PHM}
\label{sec:waveform_structure}

A gravitational-wave waveform can be decomposed as \cite{Thorne1980,Kidder2008}
\begin{equation}
h(t;\iota,\varphi_0)
=
\frac{1}{D_L}
\sum_{\ell,m}
h_{\ell m}(t)
\,{}_{-2}Y_{\ell m}(\iota,\varphi_0),
\label{eq:h_decomp}
\end{equation}
where $D_L$ is the luminosity distance and $(\iota,\varphi_0)$ describe the source orientation. 

In the effective-one-body (EOB) formalism, the $h_{\ell m}$ modes are constructed from the orbital dynamics \cite{Buonanno1999,Damour2000,Pan2011,Ossokine2020}. \texttt{SEOBNRv5EHM} \cite{SEOBNRv5EHM} and \texttt{SEOBNRv5PHM} \cite{SEOBNRv5PHM} have the same EOB backbone dynamics but differ in their treatment of eccentricity and spin precession respectively. 

\texttt{SEOBNRv5EHM} is designed to incorporate eccentricity effects, and in the circular limit its modes take a simple form \cite{SEOBNRv5EHM, Pan2011,Damour_2009}:
\begin{equation}
h^{\rm EHM}_{\ell m}(t)= A_{\ell m}(t)e^{-i \phi_{\ell m}(t)},
\end{equation}
with phase
\begin{equation}
\phi_{\ell m}(t) = m \Phi_{\rm orb}(t) + \delta_{\ell m}(t),
\label{eq:aligned_phase}
\end{equation}
where $\Phi_{\rm orb}(t)$ is the orbital phase and $\delta_{\ell m}(t)$ contains post-Newtonian and non-quasicircular corrections \cite{Hinderer_2017,Chiaramello_2020}. Each mode is therefore governed by a single dynamical phase variable,
\begin{equation}
h^{\rm EHM}_{\ell m}(t) = A_{\ell m}(t)e^{-i[m\Phi_{\rm orb}(t)+\delta_{\ell m}(t)]}.
\label{eq:v5EHM}
\end{equation}

On the other hand, \texttt{SEOBNRv5PHM} incorporates precession by constructing the waveform in a co-precessing frame and rotating to the inertial frame \cite{SEOBNRv5PHM,Ossokine2020}:
\begin{equation}
h^{\rm PHM}_{\ell m}(t) =\sum_{m'}D^\ell_{m m'}(\alpha,\beta,\gamma)\, h^{\rm copr}_{\ell m'}(t),
\label{eq:rotation}
\end{equation}
where $D^\ell_{mm'}$ are Wigner rotation matrices parameterized by the time-dependent Euler angles $(\alpha,\beta,\gamma)$ \cite{Babak_2017}. The effective spin precession, $\chi_p$, Eq.~\ref{eq:efchi}, measures the magnitude of the spin components perpendicular to the instantaneous orbital angular momentum
$\hat{\mathbf{L}}(t)$ that drive precession, and therefore sets the
time scale of the dependence of the Euler angles $(\alpha,\beta,\gamma)$
appearing in Eq.~(\ref{eq:rotation}), which in turn control the
strength of mode mixing and the resulting amplitude and phase modulations
in the inertial-frame waveform modes
\cite{Schmidt2015, Hannam2014, Ossokine2020, ramosbuades2023seobnrv5phm}.

The co-precessing modes retain the aligned-spin structure,
\begin{equation}
h^{\rm copr}_{\ell m}(t) = A^{\rm copr}_{\ell m}(t) e^{-i[m\Phi_{\rm copr}(t)+\delta_{\ell m}(t)]}.
\end{equation}

Substituting yields the inertial-frame modes,
\begin{widetext}
\begin{equation}
h^{\rm PHM}_{\ell m}(t)=\sum_{m'}D^\ell_{m m'}(\beta)A^{\rm copr}_{\ell m'}(t)
e^{-i[m\alpha+m'\gamma+m'\Phi_{\rm copr}+\delta_{\ell m'}]}.\label{eq:phm}
\end{equation}
\end{widetext}

We can see that while \texttt{SEOBNRv5EHM} contains independent modes characterized by a single orbital phase, the precession effects in the \texttt{SEOBNRv5PHM} modes contain additional phases $(\alpha,\gamma)$ and mixing controlled by $\beta$ that produce additional amplitude and phase modulations. %Notice that in the limit $\beta\to0$, SEOBNRv5PHM reduces continuously to the aligned-spin structure of SEOBNRv5EHM.

\section{The 1D wavelet Scattering Transform}
\label{Ap:1}

Following the notation used in \texttt{Kymatio}~\cite{andreux2022, Anden_2014}, the scattering transform is defined relative to the duration of the input signal. In our case, the wavelet scattering transform (WST) is applied to a discretised detector strain signal $h(t)$ of duration $T$ and sampling frequency $f_s$, so that the total number of time samples is $N = T f_s$. The Fourier frequency resolution of such a segment is set by the base frequency
\begin{equation}
\nu_0 = \frac{1}{T},
\end{equation}
i.e.\ the spacing between adjacent Fourier bins. We stress that $\nu_0$ is the resolution of the segment, not the lowest wavelet frequency: as shown below, the latter is instead controlled by the maximum scale $J$.

The scattering filter bank is composed of wavelets indexed by $\lambda = (j,q)$, where $j = 1,\dots,J$ labels the octave (dyadic scale $2^j$) and $q = 0,\dots,Q-1$ labels the position within each octave. Following the \texttt{Kymatio} \cite{andreux2022} convention, the bank is anchored at the high-frequency (Nyquist) end and extends downwards by $J$ octaves, with wavelet centre frequencies
\begin{equation}
f_{j,q} = \frac{f_s}{2}\,\, 2^{-(j-1) - q/Q},
\label{base_freq}
\end{equation}
which produces a logarithmic tiling of the frequency domain. Larger $j$ thus corresponds to a coarser dyadic scale, a lower centre frequency, and a broader temporal support.

The width parameter $\sigma_{j,q}$ sets the temporal width of the Gaussian envelope, chosen as
\begin{equation}
\sigma_{j,q}(f_{j,q},Q) = \frac{1}{f_{j,q}\left(2^{1/Q} - 2^{-1/Q}\right)},
\end{equation}
which ensures an approximately constant relative bandwidth across scales (a constant-$Q$ filter bank).

The WST is constructed using complex Morlet wavelets $\psi_{j,q}(t)$, localised in both time and frequency, given by
\begin{equation}
\psi_{j,q}(t) \approx \frac{1}{\sigma_{j,q}\sqrt{2\pi}}
\left( e^{\,i\,2\pi f_{j,q} t} \right)
e^{-t^2/(2\sigma_{j,q}^2)}, \label{eq:Wmorlet}
\end{equation}
The characteristic frequency associated with octave $j$ is 
\begin{equation}
\nu_j = \frac{f_s}{2^{\,j}}.
\label{lambda}
\end{equation}
This construction yields a constant-$Q$ filter bank spanning the physically accessible range
\begin{equation}
\frac{f_s}{2^{\,J}} \;\lesssim\; f_{j,q} \;\lesssim\; \frac{f_s}{2},
\end{equation}
where the upper bound is the Nyquist frequency and the lower bound is set by the maximum scale $J$, i.e.\ by how many octaves below Nyquist the bank extends. Increasing $j$ moves to coarser scales and lower centre frequencies, while $Q$ fixes the number of wavelets per octave and hence the frequency resolution of the tiling.

In the time domain, each wavelet has a characteristic temporal support $\Delta t_{j,q}$ inversely proportional to its bandwidth,
\begin{equation}
\Delta t_{j,q}
\;\sim\;
\frac{1}{\Delta f_{j,q}}
\;\sim\;
\frac{Q}{f_{j,q}},
\end{equation}
so that high-frequency wavelets probe rapid oscillations while low-frequency wavelets capture slower modulations of the signal~\citep{Anden_2015}. The parameter $Q$ therefore controls the time--frequency trade-off of the transform: larger $Q$ provides finer frequency resolution within each octave, while smaller $Q$ improves temporal localisation.

In practice, $J$ is chosen such that
\begin{equation}
\tau_J \;\lesssim\; T,
\end{equation}
ensuring that the averaging window does not exceed the duration of the analysed strain segment. In this work, detector data segments have duration $T = 8\,\mathrm{s}$. At a sampling frequency of $2048\,\mathrm{Hz}$, choosing $J=7$ sets $\tau_J = 1/16\,\mathrm{s}$, with a corresponding lowest wavelet frequency $\nu_J = f_s/2^{\,J} = 16\,\mathrm{Hz}$.

\section{1D-CNN vs LR and 2D-CNN}
\label{app: other classifiers}

\begin{figure}[t]
    \centering
    \includegraphics[width=0.4\textwidth]{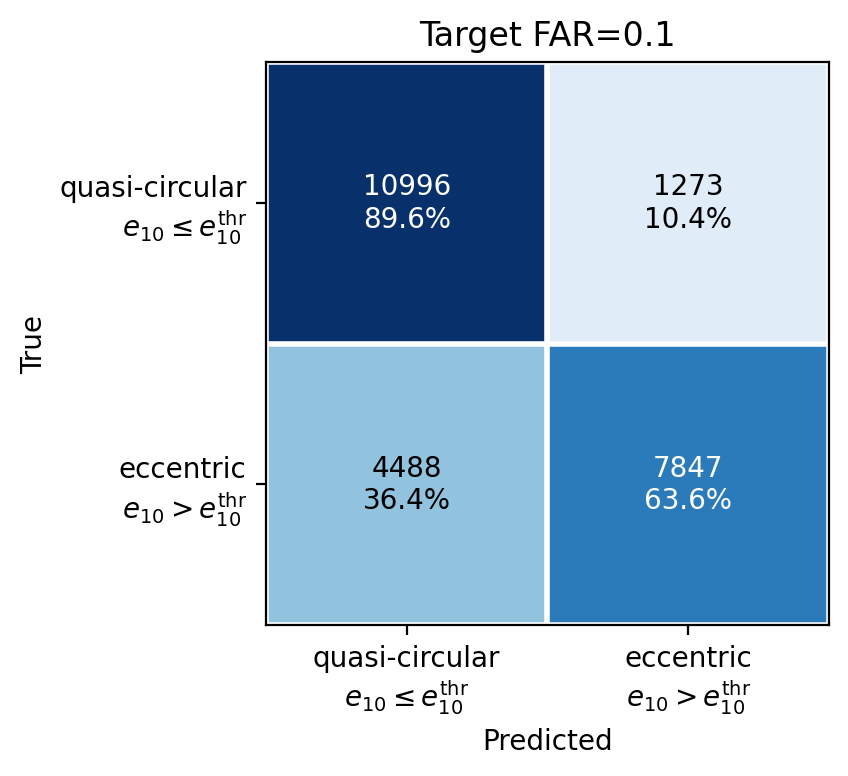}
    \caption{Confusion matrix obtained for the reference 1D-CNN model and $J=7, Q=2$ WST coefficients, tuned to a target FPR of 10\% on the validation set.}
    \label{fig:cm_j7q2}
\end{figure}

\begin{figure}[t]
    \centering
    \includegraphics[width =0.4\textwidth]{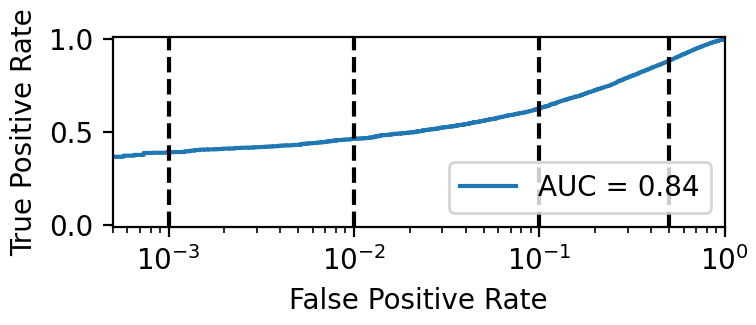}
    \caption{ROC curve and AUC on the test set, for the best fold during training of the 1D-CNN model with $J=7, Q=2$ WST coefficients.}
    \label{fig: 1dcnn_ref_test_roc}
\end{figure}

\begin{table}[t]
\centering
\caption{Performance of the alternative classifiers, the 2D-CNN and the logistic regression (LR). Performance is summarised by the threshold-independent metrics AUC (area under the ROC curve) and AP (average precision), evaluated on the test set for the best fold; the TPR is additionally reported at a target FPR of $10\%$. The different 2D-CNN set-ups are detailed in the text.}
\label{tab: model comparison LR 2D CNN}
\begin{ruledtabular}
\begin{tabular}{lccc}
\textbf{Model} & \textbf{AUC} & \textbf{AP} & \textbf{TPR$_{0.1}$} \\
\hline
\textbf{Log Reg} %& $\sim0.79^{\ast}$s 
&0.815 & 0.847 &  0.562 \\
\hline
\multicolumn{4}{l}{\textbf{2D-CNN}} \\
\hline
\textbf{Log-scale} 
%& \textbf{$\sim$6m30s} 
& \textbf{0.841} & \textbf{0.872} & \textbf{0.630} \\
Raw 
%& $\sim$3m40s 
& 0.814 & 0.849 & 0.563 \\
Large 
%& $\sim$6m20 
& 0.826 & 0.858 & 0.580 \\
\end{tabular}
\end{ruledtabular}
\end{table}
\begin{figure*}
\centering
\begin{tikzpicture}[
    block/.style={
        draw,
        rectangle,
        rounded corners,
        align=center,
        minimum height=1cm,
        minimum width=3.8cm,
        font=\footnotesize
    },
    arrow/.style={->, thick},
    node distance=0.6cm
]
% ---------- High-level architecture ----------
\node[block] (input) {Input: {\scriptsize $(N,D,C,T)$}};
\node[block, below=of input] (fund) {
Fundamental Block ($\times$5) 
};
\node[block, below=of fund] (pool) {
AvgPool2D $+$ Flatten {\scriptsize $(N,256)$}
};
\node[block, below=of pool] (dense) {
Dense Layer + ELU ($\times$3) {\scriptsize  $(N,16)$}
};
\node[block, below=of dense] (out) {
Output Layer {\scriptsize  $(N,1)$}
};
% Arrows for high-level
\draw[arrow] (input) -- (fund);
\draw[arrow] (fund) -- (pool);
\draw[arrow] (pool) -- (dense);
\draw[arrow] (dense) -- (out);
% ---------- Fundamental Block ----------
\node[block, right=6cm, yshift=-1.5cm] (sep) {SeparableConv2D};
\node[block, below=of sep] (bn) {BatchNorm};
\node[block, below=of bn] (relu) {ReLU};
\node[block, below=of relu] (pool2) {MaxPool2D};
% Arrows for fundamental block
\draw (fund) -- (sep);
\draw[arrow] (sep) -- (bn);
\draw[arrow] (bn) -- (relu);
\draw[arrow] (relu) -- (pool2);
\draw (fund.east) -- (pool2.west);
% Optional dashed box around zoom-in
\node[draw,dashed,fit=(sep)(pool2),inner sep=4pt] {};
\end{tikzpicture}
\caption{Two-component diagram of \texttt{SWT\_CNN\_2D}. Left: the high-level architecture, showing the input, the fundamental layer ($5\times$ Fundamental Block), global pooling, the dense layers, and the output. Right: a zoom-in of a single Fundamental Block (SeparableConv2D $+$ BatchNorm $+$ ReLU $+$ MaxPool). Tensor shapes are denoted $(N,D,C,T)$.}
\label{fig:swt_cnn_2d_two_component}
\end{figure*}

To set a baseline for the 1D-CNN, we apply a simpler LR classifier to the WST data. Our experiments show that an LR classifier alone is already sufficient to identify eccentricity in the data, achieving an AUC of $\sim 0.8$ and a TPR of $\sim 0.55$ at a target FPR of $10\%$.  In addition we also compare against a more complex 2D-CNN classifier that operates directly on the two-dimensional WST representation. Its input consists of the logarithmically rescaled\footnote{The raw WST coefficients span several orders of magnitude, which can adversely affect neural-network training.} WST coefficient tensors, with convolutional kernels acting along both the temporal ($T_b$) and channel ($C$) dimensions, and with the three detectors treated as separate input channels, analogous to the RGB channels in image analysis. The network is designed to resemble the \texttt{ecc\_sepcnn} model of Ref.~\cite{ravichandran2024} and is implemented in \texttt{PyTorch}~\cite{paszke2019pytorch}. Its core consists of separable convolutional layers followed by fully connected layers; a schematic overview is shown in Fig.~\ref{fig:swt_cnn_2d_two_component}. The number of channels in successive convolutional blocks increases as $32 \rightarrow 64 \rightarrow 128 \rightarrow 128 \rightarrow 256$, while the kernel sizes decrease as $5 \rightarrow 5 \rightarrow 5 \rightarrow 3 \rightarrow 3$. A dropout rate of $0.2$ is used to mitigate overfitting, and the fully connected layers reduce the feature dimension as $64 \rightarrow 32 \rightarrow 16$, followed by a single output node. With these settings, the model contains $86\,369$ trainable parameters.

In its default configuration, the best-performing fold achieves a test AUC of $0.841$ after $24$ epochs, with a TPR of $63.0\%$ at a target FPR of $10\%$. As reported in Table~\ref{tab: model comparison LR 2D CNN}, omitting the logarithmic rescaling of the WST coefficients (\emph{Raw}) degrades performance, confirming the importance of dynamic-range compression for effective feature extraction. Doubling the number of trainable parameters (\emph{Large}) yields no further gains, indicating that the baseline architecture already captures the relevant structure in the data.

Comparing all three  classifiers in Table~\ref{tab: J Q comparison 1D CNN} and \ref{tab: model comparison LR 2D CNN}, the 2D-CNN attains performance comparable to the 1D-CNN at a similar parameter count (Figures \ref{fig:cm_j7q2}, \ref{fig: 1dcnn_ref_test_roc} provide additional information on the performance of the 1D-CNN classifier), but at a substantially higher training cost. The LR and 1D-CNN are trained on a MacBook Pro (M3 chip), requiring $\sim 12$ minutes (including hyperparameter optimisation) and $\sim 10$ minutes for 10-fold cross-validation, respectively. The 2D-CNN, by contrast, is trained on the CSD3 cluster (Ampere nodes, Cambridge, UK) using one A100 GPU and 32 CPU cores, incurring significantly greater computational overhead for equivalent predictive performance. This near-equivalence suggests that preserving the two-dimensional structure of the WST coefficients confers no significant advantage: unlike conventional time--frequency representations such as $Q$-transforms, the WST coefficients do not require an image-based interpretation, and their informative content can be captured effectively by one-dimensional models at substantially lower computational cost. 

\newpage
\bibliographystyle{ieeetr}
\bibliography{references}

@article{Paul_2025,
   title={Eccentric, spinning, inspiral-merger-ringdown waveform model with higher modes for the detection and characterization of binary black holes},
   volume={111},
   ISSN={2470-0029},
   url={http://dx.doi.org/10.1103/PhysRevD.111.084074},
   DOI={10.1103/physrevd.111.084074},
   number={8},
   journal={Physical Review D},
   publisher={American Physical Society (APS)},
   author={Paul, Kaushik and Maurya, Akash and Henry, Quentin and Sharma, Kartikey and Satheesh, Pranav and Divyajyoti and Kumar, Prayush and Mishra, Chandra Kant},
   year={2025},
   month=Apr }

@article{Phukon_2025,
   title={Geometric template bank for the detection of spinning low-mass compact binaries with moderate orbital eccentricity},
   volume={111},
   ISSN={2470-0029},
   url={http://dx.doi.org/10.1103/PhysRevD.111.043040},
   DOI={10.1103/physrevd.111.043040},
   number={4},
   journal={Physical Review D},
   publisher={American Physical Society (APS)},
   author={Phukon, Khun Sang and Schmidt, Patricia and Pratten, Geraint},
   year={2025},
   month=Feb }

@misc{wst-eccentricity,
  author       = {P. Canizares and S. J. Staelens and I. Romero-Shaw},
  title        = {{WST-Eccentricity}},
  year         = {2026},
  howpublished = {\url{https://github.com/Priscilla-/wst-eccentricity}},
  note         = {Accessed: 2026-07-07}
}

@article{scikit-learn,
  title={Scikit-learn: Machine Learning in {P}ython},
  author={Pedregosa, F. and Varoquaux, G. and Gramfort, A. and Michel, V.
          and Thirion, B. and Grisel, O. and Blondel, M. and Prettenhofer, P.
          and Weiss, R. and Dubourg, V. and Vanderplas, J. and Passos, A. and
          Cournapeau, D. and Brucher, M. and Perrot, M. and Duchesnay, E.},
  journal={Journal of Machine Learning Research},
  volume={12},
  pages={2825--2830},
  year={2011}
}

@misc{tibrewal2026,
      title={Misinterpreting Spin Precession as Orbital Eccentricity in Gravitational-Wave Signals}, 
      author={Snehal Tibrewal and Aaron Zimmerman and Jacob Lange and Deirdre Shoemaker},
      year={2026},
      eprint={2601.02260},
      archivePrefix={arXiv},
      primaryClass={gr-qc},
      url={https://arxiv.org/abs/2601.02260}, 
}

@misc{tang2026,
      title={Impact of Higher-Order Modes on Eccentricity Measurement in Binary Black Hole Gravitational Waves}, 
      author={Honglue Tang and Jinzhao Yang and Baoxiang Wang and Tao Yang},
      year={2026},
      eprint={2602.04642},
      archivePrefix={arXiv},
      primaryClass={gr-qc},
      url={https://arxiv.org/abs/2602.04642}, 
}

@misc{mallat2012,
  author       = {Mallat, Stéphane},
  title        = {Group Invariant Scattering},
  year         = {2012},
  eprint       = {1101.2286},
  archivePrefix= {arXiv},
  primaryClass = {math.FA},
  url          = {https://arxiv.org/abs/1101.2286}
}

@article{LIGOScientific:2026wfs,
    author = "Abac, None and others",
    collaboration = "LIGO Scientific, VIRGO, KAGRA",
    title = "{GWTC-5.0: Observations from the Second Part of the Fourth LIGO-Virgo-KAGRA Observing Run and Updates to the Gravitational-Wave Transient Catalog}",
    eprint = "2605.27225",
    archivePrefix = "arXiv",
    primaryClass = "gr-qc",
    reportNumber = "LIGO-P2600152",
    month = "5",
    year = "2026"
}

@article{LIGOScientific:2026ctl,
    author = "Abac, None and others",
    collaboration = "LIGO Scientific, VIRGO, KAGRA",
    title = "{GWTC-5.0: Population Properties of Merging Compact Binaries}",
    eprint = "2605.27226",
    archivePrefix = "arXiv",
    primaryClass = "astro-ph.HE",
    reportNumber = "LIGO-P2600045",
    month = "5",
    year = "2026"
}

@article{bilby1,
	adsurl = {https://ui.adsabs.harvard.edu/abs/2019ApJS..241...27A},
	archiveprefix = {arXiv},
	author = {{Ashton}, Gregory and {H{\"u}bner}, Moritz and {Lasky}, Paul D. and {Talbot}, Colm and {Ackley}, Kendall and {Biscoveanu}, Sylvia and {Chu}, Qi and {Divakarla}, Atul and {Easter}, Paul J. and {Goncharov}, Boris and {Hernandez Vivanco}, Francisco and {Harms}, Jan and {Lower}, Marcus E. and {Meadors}, Grant D. and {Melchor}, Denyz and {Payne}, Ethan and {Pitkin}, Matthew D. and {Powell}, Jade and {Sarin}, Nikhil and {Smith}, Rory J.~E. and {Thrane}, Eric},
	doi = {10.3847/1538-4365/ab06fc},
	eid = {27},
	eprint = {1811.02042},
	journal = {Astrophys. J. Supp. S.},
	month = apr,
	number = {2},
	pages = {27},
	primaryclass = {astro-ph.IM},
	title = {{BILBY: A User-friendly Bayesian Inference Library for Gravitational-wave Astronomy}},
	volume = {241},
	year = 2019}

@article{bilby2,
	adsurl = {https://ui.adsabs.harvard.edu/abs/2020MNRAS.499.3295R},
	archiveprefix = {arXiv},
	author = {{Romero-Shaw}, I.~M. and others},
	doi = {10.1093/mnras/staa2850},
	eprint = {2006.00714},
	journal = {Mon. Not. R. Astron. Soc.},
	month = dec,
	number = {3},
	pages = {3295-3319},
	primaryclass = {astro-ph.IM},
	title = {{Bayesian inference for compact binary coalescences with BILBY: validation and application to the first LIGO-Virgo gravitational-wave transient catalogue}},
	volume = {499},
	year = 2020}

@misc{bruna2011,
  author       = {Bruna, Joan and Mallat, Stéphane},
  title        = {Classification with Invariant Scattering Representations},
  year         = {2011},
  eprint       = {1112.1120},
  archivePrefix= {arXiv},
  primaryClass = {cs.CV},
  url          = {https://arxiv.org/abs/1112.1120}
}

@misc{andreux2022,
  author       = {Andreux, Mathieu and Angles, Tomás and Exarchakis, Georgios and Leonarduzzi, Roberto and Rochette, Gaspar and Thiry, Louis and Zarka, John and Mallat, Stéphane and Andén, Joakim and Belilovsky, Eugene and Bruna, Joan and Lostanlen, Vincent and Chaudhary, Muawiz and Hirn, Matthew J. and Oyallon, Edouard and Zhang, Sixin and Cella, Carmine and Eickenberg, Michael},
  title        = {Kymatio: Scattering Transforms in Python},
  year         = {2022},
  eprint       = {1812.11214},
  archivePrefix= {arXiv},
  primaryClass = {cs.LG},
  url          = {https://arxiv.org/abs/1812.11214}
}

@misc{patterson2024,
  author       = {Patterson, Ben G. and Tomson, Sharon Mary and Fairhurst, Stephen},
  title        = {Identifying Eccentricity in Binary Black Hole Mergers Using a Harmonic Decomposition of the Gravitational Waveform},
  year         = {2024},
  eprint       = {2411.04187},
  archivePrefix= {arXiv},
  primaryClass = {gr-qc},
  url          = {https://arxiv.org/abs/2411.04187}
}

@article{Damour_2009,
   title={Improved resummation of post-Newtonian multipolar waveforms from circularized compact binaries},
   volume={79},
   ISSN={1550-2368},
   url={http://dx.doi.org/10.1103/PhysRevD.79.064004},
   DOI={10.1103/physrevd.79.064004},
   number={6},
   journal={Physical Review D},
   publisher={American Physical Society (APS)},
   author={Damour, Thibault and Iyer, Bala R. and Nagar, Alessandro},
   year={2009},
   month=mar }

@article{Shaikh_2023,
  author       = {Shaikh, Md Arif and Varma, Vijay and Pfeiffer, Harald P. and Ramos-Buades, Antoni and van de Meent, Maarten},
  title        = {Defining eccentricity for gravitational wave astronomy},
  journal      = {Physical Review D},
  volume       = {108},
  number       = {10},
  pages        = {104007},
  year         = {2023},
  month        = {nov},
  doi          = {10.1103/PhysRevD.108.104007},
  url          = {https://doi.org/10.1103/PhysRevD.108.104007}
}

@misc{islam2024b,
  author       = {Islam, Tousif and Khanna, Gaurav and Field, Scott E.},
  title        = {Adding Higher-order Spherical Harmonics in Non-spinning Eccentric Binary Black Hole Merger Waveform Models},
  year         = {2024},
  eprint       = {2408.02762},
  archivePrefix= {arXiv},
  primaryClass = {gr-qc},
  url          = {https://arxiv.org/abs/2408.02762}
}

@article{Vijaykumar_2024,
  author       = {Vijaykumar, Aditya and Hanselman, Alexandra G. and Zevin, Michael},
  title        = {Consistent Eccentricities for Gravitational-wave Astronomy: Resolving Discrepancies between Astrophysical Simulations and Waveform Models},
  journal      = {The Astrophysical Journal},
  volume       = {969},
  number       = {2},
  pages        = {132},
  year         = {2024},
  month        = {jul},
  doi          = {10.3847/1538-4357/ad4455},
  url          = {https://doi.org/10.3847/1538-4357/ad4455}
}

@misc{ravichandran2024,
  author       = {Ravichandran, Adhrit and Vijaykumar, Aditya and Kapadia, Shasvath J. and Kumar, Prayush},
  title        = {Rapid Identification and Classification of Eccentric Gravitational Wave Inspirals with Machine Learning},
  year         = {2024},
  eprint       = {2302.00666},
  archivePrefix= {arXiv},
  primaryClass = {gr-qc},
  url          = {https://arxiv.org/abs/2302.00666}
}

@article{Romero_Shaw_2023,
  author       = {Romero-Shaw, Isobel M. and Gerosa, Davide and Loutrel, Nicholas},
  title        = {Eccentricity or spin precession? Distinguishing subdominant effects in gravitational-wave data},
  journal      = {Monthly Notices of the Royal Astronomical Society},
  volume       = {519},
  number       = {4},
  pages        = {5352--5357},
  year         = {2023},
  month        = {jan},
  doi          = {10.1093/mnras/stad031},
  url          = {https://doi.org/10.1093/mnras/stad031}
}

@article{P_64,
  author       = {Peters, P. C.},
  title        = {Gravitational Radiation and the Motion of Two Point Masses},
  journal      = {Physical Review},
  volume       = {136},
  pages        = {B1224--B1232},
  year         = {1964},
  month        = {nov},
  doi          = {10.1103/PhysRev.136.B1224},
  url          = {https://doi.org/10.1103/PhysRev.136.B1224}
}

@article{gamboa2025,
  author       = {Gamboa, Aldo and Buonanno, Alessandra and Enficiaud, Raffi and Khalil, Mohammed and Ramos-Buades, Antoni and Pompili, Lorenzo and Estellés, Héctor and Boyle, Michael and Kidder, Lawrence E. and Pfeiffer, Harald P. and Rüter, Hannes R. and Scheel, Mark A.},
  title        = {Accurate waveforms for eccentric, aligned-spin binary black holes: The multipolar effective-one-body model SEOBNRv5EHM},
  journal      = {Physical Review D},
  volume       = {112},
  number       = {4},
  pages        = {044038},
  year         = {2025},
  month        = {aug},
  doi          = {10.1103/jxrc-z298},
  url          = {https://doi.org/10.1103/jxrc-z298}
}

@misc{bruna2012,
  author       = {Bruna, Joan and Mallat, Stéphane},
  title        = {Invariant Scattering Convolution Networks},
  year         = {2012},
  eprint       = {1203.1513},
  archivePrefix= {arXiv},
  primaryClass = {cs.CV},
  url          = {https://arxiv.org/abs/1203.1513}
}

@inproceedings{Anden_2015,
  author       = {Andén, Joakim and Lostanlen, Vincent and Mallat, Stéphane},
  title        = {Joint Time-Frequency Scattering for Audio Classification},
  booktitle    = {2015 IEEE 25th International Workshop on Machine Learning for Signal Processing (MLSP)},
  publisher    = {IEEE},
  pages        = {1--6},
  year         = {2015},
  month        = {sep},
  doi          = {10.1109/MLSP.2015.7324385},
  url          = {https://doi.org/10.1109/MLSP.2015.7324385}
}

@article{Anden_2014,
   title={Deep Scattering Spectrum},
   volume={62},
   ISSN={1941-0476},
   url={http://dx.doi.org/10.1109/TSP.2014.2326991},
   DOI={10.1109/tsp.2014.2326991},
   number={16},
   journal={IEEE Transactions on Signal Processing},
   publisher={Institute of Electrical and Electronics Engineers (IEEE)},
   author={Anden, Joakim and Mallat, Stephane},
   year={2014},
   month=Aug, pages={4114–4128} }

@article{Cheng_2024,
  author       = {Cheng, Sihao and Morel, Rudy and Allys, Erwan and Ménard, Brice and Mallat, Stéphane},
  title        = {Scattering spectra models for physics},
  journal      = {PNAS Nexus},
  volume       = {3},
  number       = {4},
  pages        = {pgae103},
  year         = {2024},
  month        = {mar},
  doi          = {10.1093/pnasnexus/pgae103},
  url          = {https://doi.org/10.1093/pnasnexus/pgae103}
}

@misc{paszke2019pytorch,
  author       = {Paszke, Adam and Gross, Sam and Massa, Francisco and Lerer, Adam and Bradbury, James and Chanan, Gregory and Killeen, Trevor and Lin, Zeming and Gimelshein, Natalia and Antiga, Luca and Desmaison, Alban and Köpf, Andreas and Yang, Edward and DeVito, Zach and Raison, Martin and Tejani, Alykhan and Chilamkurthy, Sasank and Steiner, Benoit and Fang, Lu and Bai, Junjie and Chintala, Soumith},
  title        = {PyTorch: An Imperative Style, High-Performance Deep Learning Library},
  year         = {2019},
  eprint       = {1912.01703},
  archivePrefix= {arXiv},
  primaryClass = {cs.LG},
  url          = {https://arxiv.org/abs/1912.01703}
}

@misc{ramosbuades2023seobnrv5phm,
  author       = {Ramos-Buades, Antoni and Buonanno, Alessandra and Estellés, Héctor and Khalil, Mohammed and Mihaylov, Deyan P. and Ossokine, Serguei and Pompili, Lorenzo and Shiferaw, Mahlet},
  title        = {SEOBNRv5PHM: Next Generation of Accurate and Efficient Multipolar Precessing-spin Effective-one-body Waveforms for Binary Black Holes},
  year         = {2023},
  eprint       = {2303.18046},
  archivePrefix= {arXiv},
  primaryClass = {gr-qc},
  url          = {https://arxiv.org/abs/2303.18046}
}

@article{Thorne1980,
  author       = {Thorne, Kip S.},
  title        = {Multipole expansions of gravitational radiation},
  journal      = {Reviews of Modern Physics},
  volume       = {52},
  number       = {2},
  pages        = {299--339},
  year         = {1980},
  doi          = {10.1103/RevModPhys.52.299},
  url          = {https://doi.org/10.1103/RevModPhys.52.299}
}

@article{Kidder2008,
  author       = {Kidder, Lawrence E.},
  title        = {Using full information when computing modes of post-Newtonian waveforms from inspiralling compact binaries in circular orbit},
  journal      = {Physical Review D},
  volume       = {77},
  number       = {4},
  pages        = {044016},
  year         = {2008},
  doi          = {10.1103/PhysRevD.77.044016},
  eprint       = {0710.0614},
  archivePrefix= {arXiv},
  primaryClass = {gr-qc},
  url          = {https://doi.org/10.1103/PhysRevD.77.044016}
}

@article{Buonanno1999,
  author       = {Buonanno, Alessandra and Damour, Thibault},
  title        = {Effective one-body approach to general relativistic two-body dynamics},
  journal      = {Physical Review D},
  volume       = {59},
  number       = {8},
  pages        = {084006},
  year         = {1999},
  doi          = {10.1103/PhysRevD.59.084006},
  eprint       = {gr-qc/9811091},
  archivePrefix= {arXiv},
  url          = {https://doi.org/10.1103/PhysRevD.59.084006}
}

@article{Damour2000,
  author       = {Damour, Thibault and Jaranowski, Piotr and Sch{\"a}fer, Gerhard},
  title        = {On the determination of the last stable orbit for circular general relativistic binaries at the third post-Newtonian approximation},
  journal      = {Physical Review D},
  volume       = {62},
  number       = {8},
  pages        = {084011},
  year         = {2000},
  doi          = {10.1103/PhysRevD.62.084011},
  eprint       = {gr-qc/0005034},
  archivePrefix= {arXiv},
  url          = {https://doi.org/10.1103/PhysRevD.62.084011}
}

@article{Pan2011,
  author       = {Pan, Yi and Buonanno, Alessandra and Boyle, Michael and Buchman, Luisa T. and Kidder, Lawrence E. and Pfeiffer, Harald P. and Scheel, Mark A.},
  title        = {Inspiral-merger-ringdown waveforms of spinning, precessing black-hole binaries in the effective-one-body formalism},
  journal      = {Physical Review D},
  volume       = {84},
  number       = {12},
  pages        = {124052},
  year         = {2011},
  doi          = {10.1103/PhysRevD.84.124052},
  eprint       = {1106.1021},
  archivePrefix= {arXiv},
  primaryClass = {gr-qc},
  url          = {https://doi.org/10.1103/PhysRevD.84.124052}
}

@article{Ossokine2020,
  author       = {Ossokine, Serguei and others},
  title        = {Multipolar effective-one-body waveforms for precessing binary black holes: Construction and validation},
  journal      = {Physical Review D},
  volume       = {102},
  number       = {4},
  pages        = {044055},
  year         = {2020},
  doi          = {10.1103/PhysRevD.102.044055},
  eprint       = {2004.09442},
  archivePrefix= {arXiv},
  primaryClass = {gr-qc},
  url          = {https://doi.org/10.1103/PhysRevD.102.044055}
}

@ARTICLE{Stegmann25,
       author = {{Stegmann}, Jakob and {Antonini}, Fabio and {Olejak}, Aleksandra and {Biscoveanu}, Sylvia and {Raymond}, Vivien and {Rinaldi}, Stefano and {Flanagan}, Beth},
        title = "{In-plane Black-hole Spin Measurements Suggest Most Gravitational-wave Mergers Form in Triples}",
      journal = {arXiv e-prints},
         year = 2025,
        month = dec,
          eid = {arXiv:2512.15873},
        pages = {arXiv:2512.15873},
          doi = {10.48550/arXiv.2512.15873},
archivePrefix = {arXiv},
       eprint = {2512.15873},
 primaryClass = {astro-ph.HE},
       adsurl = {https://ui.adsabs.harvard.edu/abs/2025arXiv251215873S}
}

@ARTICLE{Antonini17,
       author = {{Antonini}, Fabio and {Toonen}, Silvia and {Hamers}, Adrian S.},
        title = "{Binary Black Hole Mergers from Field Triples: Properties, Rates, and the Impact of Stellar Evolution}",
      journal = {\apj},
         year = 2017,
        month = jun,
       volume = {841},
       number = {2},
          eid = {77},
        pages = {77},
          doi = {10.3847/1538-4357/aa6f5e},
archivePrefix = {arXiv},
       eprint = {1703.06614},
 primaryClass = {astro-ph.GA},
       adsurl = {https://ui.adsabs.harvard.edu/abs/2017ApJ...841...77A}
}

@ARTICLE{IRS22,
       author = {{Romero-Shaw}, Isobel and {Lasky}, Paul D. and {Thrane}, Eric},
        title = "{Four Eccentric Mergers Increase the Evidence that LIGO-Virgo-KAGRA's Binary Black Holes Form Dynamically}",
      journal = {\apj},
         year = 2022,
        month = dec,
       volume = {940},
       number = {2},
          eid = {171},
        pages = {171},
          doi = {10.3847/1538-4357/ac9798},
archivePrefix = {arXiv},
       eprint = {2206.14695},
 primaryClass = {astro-ph.HE},
       adsurl = {https://ui.adsabs.harvard.edu/abs/2022ApJ...940..171R}
}

@ARTICLE{Morras25,
       author = {{Morras}, Gonzalo and {Pratten}, Geraint and {Schmidt}, Patricia},
        title = "{Orbital eccentricity in a neutron star - black hole binary}",
      journal = {arXiv e-prints},
         year = 2025,
        month = mar,
          eid = {arXiv:2503.15393},
        pages = {arXiv:2503.15393},
          doi = {10.48550/arXiv.2503.15393},
archivePrefix = {arXiv},
       eprint = {2503.15393},
 primaryClass = {astro-ph.HE},
       adsurl = {https://ui.adsabs.harvard.edu/abs/2025arXiv250315393M}
}

@ARTICLE{Planas25,
       author = {{Planas}, Maria de Lluc and {Ramos-Buades}, Antoni and {Garc{\'\i}a-Quir{\'o}s}, Cecilio and {Estell{\'e}s}, H{\'e}ctor and {Husa}, Sascha and {Haney}, Maria},
        title = "{Reanalysis of binary black hole gravitational wave events for orbital eccentricity signatures}",
      journal = {\prd},
         year = 2025,
        month = dec,
       volume = {112},
       number = {12},
          eid = {123004},
        pages = {123004},
          doi = {10.1103/cv75-y8dr},
archivePrefix = {arXiv},
       eprint = {2504.15833},
 primaryClass = {gr-qc},
       adsurl = {https://ui.adsabs.harvard.edu/abs/2025PhRvD.112l3004P}
}

@ARTICLE{Xu25,
       author = {{Xu}, Yumeng and {Valencia}, Jorge and {Estell{\'e}s Estrella}, H{\'e}ctor and {Ramos Buades}, Antoni and {Husa}, Sascha and {Rossell{\'o}-Sastre}, Maria and {Llobera Querol}, Joan and {Ramis Vidal}, Felip and {de Lluc Planas Llompart}, Maria and {Colleoni}, Marta and {Hamilton}, Eleanor and {Montava Agudo}, Arnau and {Y{\'e}bana Carrilero}, Jes{\'u}s and {Heffernan}, Anna},
        title = "{Parameter estimation for the GWTC-4.0 catalog with phenomenological waveform models that include orbital eccentricity and an updated description of spin precession}",
      journal = {arXiv e-prints},
         year = 2025,
        month = dec,
          eid = {arXiv:2512.19513},
        pages = {arXiv:2512.19513},
          doi = {10.48550/arXiv.2512.19513},
archivePrefix = {arXiv},
       eprint = {2512.19513},
 primaryClass = {gr-qc},
       adsurl = {https://ui.adsabs.harvard.edu/abs/2025arXiv251219513X}
}

@ARTICLE{Tibrewal26,
       author = {{Tibrewal}, Snehal and {Zimmerman}, Aaron and {Lange}, Jacob and {Shoemaker}, Deirdre},
        title = "{Misinterpreting Spin Precession as Orbital Eccentricity in Gravitational-Wave Signals}",
      journal = {arXiv e-prints},
         year = 2026,
        month = jan,
          eid = {arXiv:2601.02260},
        pages = {arXiv:2601.02260},
          doi = {10.48550/arXiv.2601.02260},
archivePrefix = {arXiv},
       eprint = {2601.02260},
 primaryClass = {gr-qc},
       adsurl = {https://ui.adsabs.harvard.edu/abs/2026arXiv260102260T}
}

@ARTICLE{2022PhRvD.106j4017P,
       author = {{Payne}, Ethan and {Hourihane}, Sophie and {Golomb}, Jacob and {Udall}, Rhiannon and {Davis}, Derek and {Chatziioannou}, Katerina},
        title = "{Curious case of GW200129: Interplay between spin-precession inference and data-quality issues}",
      journal = {\prd},
         year = 2022,
        month = nov,
       volume = {106},
       number = {10},
          eid = {104017},
        pages = {104017},
          doi = {10.1103/PhysRevD.106.104017},
archivePrefix = {arXiv},
       eprint = {2206.11932},
 primaryClass = {gr-qc},
       adsurl = {https://ui.adsabs.harvard.edu/abs/2022PhRvD.106j4017P}
}

@ARTICLE{2025PhRvD.112f3052R,
       author = {{Romero-Shaw}, Isobel and {Stegmann}, Jakob and {Tagawa}, Hiromichi and {Gerosa}, Davide and {Samsing}, Johan and {Gupte}, Nihar and {Green}, Stephen R.},
        title = "{GW200208\_222617 as an eccentric black-hole binary merger: Properties and astrophysical implications}",
      journal = {\prd},
         year = 2025,
        month = sep,
       volume = {112},
       number = {6},
          eid = {063052},
        pages = {063052},
          doi = {10.1103/jj7m-x66y},
archivePrefix = {arXiv},
       eprint = {2506.17105},
 primaryClass = {astro-ph.HE},
       adsurl = {https://ui.adsabs.harvard.edu/abs/2025PhRvD.112f3052R}
}

@ARTICLE{2021ApJ...921L..43Z,
       author = {{Zevin}, Michael and {Romero-Shaw}, Isobel M. and {Kremer}, Kyle and {Thrane}, Eric and {Lasky}, Paul D.},
        title = "{Implications of Eccentric Observations on Binary Black Hole Formation Channels}",
      journal = {The Astrophysical Journal Letters},
         year = 2021,
        month = nov,
       volume = {921},
       number = {2},
          eid = {L43},
        pages = {L43},
          doi = {10.3847/2041-8213/ac32dc},
archivePrefix = {arXiv},
       eprint = {2106.09042},
 primaryClass = {astro-ph.HE},
       adsurl = {https://ui.adsabs.harvard.edu/abs/2021ApJ...921L..43Z}
}

@ARTICLE{2025PhRvD.111h4044L,
       author = {{Licciardi}, Alessandro and {Carbone}, Davide and {Rondoni}, Lamberto and {Nagar}, Alessandro},
        title = "{Wavelet scattering transform for gravitational wave analysis: An application to glitch characterization}",
      journal = {\prd},
         year = 2025,
        month = apr,
       volume = {111},
       number = {8},
          eid = {084044},
        pages = {084044},
          doi = {10.1103/PhysRevD.111.084044},
archivePrefix = {arXiv},
       eprint = {2411.19122},
 primaryClass = {gr-qc},
       adsurl = {https://ui.adsabs.harvard.edu/abs/2025PhRvD.111h4044L}
}

@ARTICLE{2025arXiv251216289R,
       author = {{Romero-Shaw}, Isobel and {Stegmann}, Jakob and {Morras}, Gonzalo and {Dorozsmai}, Andris and {Zevin}, Michael},
        title = "{Astrophysical Implications of Eccentricity in Gravitational Waves from Neutron Star-Black Hole Binaries}",
      journal = {arXiv e-prints},
         year = 2025,
        month = dec,
          eid = {arXiv:2512.16289},
        pages = {arXiv:2512.16289},
          doi = {10.48550/arXiv.2512.16289},
archivePrefix = {arXiv},
       eprint = {2512.16289},
 primaryClass = {astro-ph.HE},
       adsurl = {https://ui.adsabs.harvard.edu/abs/2025arXiv251216289R}
}

@ARTICLE{2010PhRvD..81b4007B,
       author = {{Brown}, Duncan A. and {Zimmerman}, Peter J.},
        title = "{Effect of eccentricity on searches for gravitational waves from coalescing compact binaries in ground-based detectors}",
      journal = {\prd},
         year = 2010,
        month = jan,
       volume = {81},
       number = {2},
          eid = {024007},
        pages = {024007},
          doi = {10.1103/PhysRevD.81.024007},
archivePrefix = {arXiv},
       eprint = {0909.0066},
 primaryClass = {gr-qc},
       adsurl = {https://ui.adsabs.harvard.edu/abs/2010PhRvD..81b4007B}
}

@ARTICLE{2024PhRvD.110d4013G,
       author = {{Gadre}, Bhooshan and {Soni}, Kanchan and {Tiwari}, Shubhanshu and {Ramos-Buades}, Antoni and {Haney}, Maria and {Mitra}, Sanjit},
        title = "{Detectability of eccentric binary black holes with matched filtering and unmodeled pipelines during the third observing run of LIGO-Virgo-KAGRA}",
      journal = {\prd},
         year = 2024,
        month = aug,
       volume = {110},
       number = {4},
          eid = {044013},
        pages = {044013},
          doi = {10.1103/PhysRevD.110.044013},
archivePrefix = {arXiv},
       eprint = {2405.04186},
 primaryClass = {gr-qc},
       adsurl = {https://ui.adsabs.harvard.edu/abs/2024PhRvD.110d4013G}
}

@ARTICLE{2020PhRvD.102d3005R,
       author = {{Ramos-Buades}, Antoni and {Tiwari}, Shubhanshu and {Haney}, Maria and {Husa}, Sascha},
        title = "{Impact of eccentricity on the gravitational-wave searches for binary black holes: High mass case}",
      journal = {\prd},
         year = 2020,
        month = aug,
       volume = {102},
       number = {4},
          eid = {043005},
        pages = {043005},
          doi = {10.1103/PhysRevD.102.043005},
archivePrefix = {arXiv},
       eprint = {2005.14016},
 primaryClass = {gr-qc},
       adsurl = {https://ui.adsabs.harvard.edu/abs/2020PhRvD.102d3005R}
}

@ARTICLE{2019MNRAS.490.5210R,
       author = {{Romero-Shaw}, Isobel M. and {Lasky}, Paul D. and {Thrane}, Eric},
        title = "{Searching for eccentricity: signatures of dynamical formation in the first gravitational-wave transient catalogue of LIGO and Virgo}",
      journal = {Monthly Notices of the Royal Astronomical Society},
         year = 2019,
        month = dec,
       volume = {490},
       number = {4},
        pages = {5210-5216},
          doi = {10.1093/mnras/stz2996},
archivePrefix = {arXiv},
       eprint = {1909.05466},
 primaryClass = {astro-ph.HE},
       adsurl = {https://ui.adsabs.harvard.edu/abs/2019MNRAS.490.5210R}
}

@ARTICLE{Morras,
       author = {{Morras}, Gonzalo and {Pratten}, Geraint and {Schmidt}, Patricia},
        title = "{Improved post-Newtonian waveform model for inspiralling precessing-eccentric compact binaries}",
      journal = {\prd},
         year = 2025,
        month = apr,
       volume = {111},
       number = {8},
          eid = {084052},
        pages = {084052},
          doi = {10.1103/PhysRevD.111.084052},
archivePrefix = {arXiv},
       eprint = {2502.03929},
 primaryClass = {gr-qc},
       adsurl = {https://ui.adsabs.harvard.edu/abs/2025PhRvD.111h4052M}
}

@ARTICLE{2025PhRvD.112l1503A,
       author = {{Albanesi}, Simone and {Gamba}, Rossella and {Bernuzzi}, Sebastiano and {Fontbut{\'e}}, Joan and {Gonzalez}, Alejandra and {Nagar}, Alessandro},
        title = "{Effective-one-body modeling for generic compact binaries with arbitrary orbits}",
      journal = {\prd},
         year = 2025,
        month = dec,
       volume = {112},
       number = {12},
          eid = {L121503},
        pages = {L121503},
          doi = {10.1103/3snf-w1x7},
archivePrefix = {arXiv},
       eprint = {2503.14580},
 primaryClass = {gr-qc},
       adsurl = {https://ui.adsabs.harvard.edu/abs/2025PhRvD.112l1503A}
}

@ARTICLE{Divyajyoti25,
       author = {{Divyajyoti} and {Romero-Shaw}, Isobel M. and {Prasad}, Vaishak and {Paul}, Kaushik and {Kant Mishra}, Chandra and {Kumar}, Prayush and {Maurya}, Akash and {Boyle}, Michael and {Kidder}, Lawrence E. and {Pfeiffer}, Harald P. and {Scheel}, Mark A.},
        title = "{Biased parameter inference of eccentric, spin-precessing binary black holes}",
      journal = {arXiv e-prints},
         year = 2025,
        month = oct,
          eid = {arXiv:2510.04332},
        pages = {arXiv:2510.04332},
          doi = {10.48550/arXiv.2510.04332},
archivePrefix = {arXiv},
       eprint = {2510.04332},
 primaryClass = {gr-qc},
       adsurl = {https://ui.adsabs.harvard.edu/abs/2025arXiv251004332D}
}

@ARTICLE{Divyajyoti24,
       author = {{Divyajyoti}, Sumit, Kumar and {Tibrewal}, Snehal and {Romero-Shaw}, Isobel M. and {Mishra}, Chandra Kant},
        title = "{Blind spots and biases: The dangers of ignoring eccentricity in gravitational-wave signals from binary black holes}",
      journal = {\prd},
         year = 2024,
        month = feb,
       volume = {109},
       number = {4},
          eid = {043037},
        pages = {043037},
          doi = {10.1103/PhysRevD.109.043037},
archivePrefix = {arXiv},
       eprint = {2309.16638},
 primaryClass = {gr-qc},
       adsurl = {https://ui.adsabs.harvard.edu/abs/2024PhRvD.109d3037D}
}

@ARTICLE{Gupte24,
       author = {{Gupte}, Nihar and {Ramos-Buades}, Antoni and {Buonanno}, Alessandra and {Gair}, Jonathan and {Miller}, M. Coleman and {Dax}, Maximilian and {Green}, Stephen R. and {P{\"u}rrer}, Michael and {Wildberger}, Jonas and {Macke}, Jakob and {Romero-Shaw}, Isobel M. and {Sch{\"o}lkopf}, Bernhard},
        title = "{Evidence for eccentricity in the population of binary black holes observed by LIGO-Virgo-KAGRA}",
      journal = {arXiv e-prints},
         year = 2024,
        month = apr,
          eid = {arXiv:2404.14286},
        pages = {arXiv:2404.14286},
          doi = {10.48550/arXiv.2404.14286},
archivePrefix = {arXiv},
       eprint = {2404.14286},
 primaryClass = {gr-qc},
       adsurl = {https://ui.adsabs.harvard.edu/abs/2024arXiv240414286G}
}

@ARTICLE{Samsing18,
       author = {{Samsing}, Johan},
        title = "{Eccentric black hole mergers forming in globular clusters}",
      journal = {\prd},
         year = 2018,
        month = may,
       volume = {97},
       number = {10},
          eid = {103014},
        pages = {103014},
          doi = {10.1103/PhysRevD.97.103014},
archivePrefix = {arXiv},
       eprint = {1711.07452},
 primaryClass = {astro-ph.HE},
       adsurl = {https://ui.adsabs.harvard.edu/abs/2018PhRvD..97j3014S}
}

@article{Babak_2017,
   title={Validating the effective-one-body model of spinning, precessing binary black holes against numerical relativity},
   volume={95},
   ISSN={2470-0029},
   url={http://dx.doi.org/10.1103/PhysRevD.95.024010},
   DOI={10.1103/physrevd.95.024010},
   number={2},
   journal={Physical Review D},
   publisher={American Physical Society (APS)},
   author={Babak, Stanislav and Taracchini, Andrea and Buonanno, Alessandra},
   year={2017},
   month=jan }

@ARTICLE{2025arXiv250904637A,
       author = {{Antonini}, Fabio and {Romero-Shaw}, Isobel and {Callister}, Thomas and {Dosopoulou}, Fani and {Chattopadhyay}, Debatri and {Gieles}, Mark and {Mapelli}, Michela},
        title = "{Gravitational waves reveal the pair-instability mass gap and constrain nuclear burning in massive stars}",
      journal = {arXiv e-prints},
         year = 2025,
        month = sep,
          eid = {arXiv:2509.04637},
        pages = {arXiv:2509.04637},
          doi = {10.48550/arXiv.2509.04637},
archivePrefix = {arXiv},
       eprint = {2509.04637},
 primaryClass = {astro-ph.HE},
       adsurl = {https://ui.adsabs.harvard.edu/abs/2025arXiv250904637A}
}

@ARTICLE{2025arXiv251105316T,
       author = {{Tong}, Hui and {Callister}, Thomas A. and {Fishbach}, Maya and {Thrane}, Eric and {Antonini}, Fabio and {Stevenson}, Simon and {Romero-Shaw}, Isobel M. and {Dosopoulou}, Fani},
        title = "{A subpopulation of low-mass, spinning black holes: signatures of dynamical assembly}",
      journal = {arXiv e-prints},
         year = 2025,
        month = nov,
          eid = {arXiv:2511.05316},
        pages = {arXiv:2511.05316},
          doi = {10.48550/arXiv.2511.05316},
archivePrefix = {arXiv},
       eprint = {2511.05316},
 primaryClass = {astro-ph.HE},
       adsurl = {https://ui.adsabs.harvard.edu/abs/2025arXiv251105316T}
}

@article{SEOBNRv5PHM,
  author       = {Pompili, Lorenzo and others},
  title        = {SEOBNRv5PHM: A Next-generation Effective-one-body Model for Precessing Binary Black Holes with Higher Harmonics},
  journal      = {arXiv e-prints},
  pages        = {arXiv:2303.18046},
  year         = {2023},
  eprint       = {2303.18046},
  archivePrefix= {arXiv},
  primaryClass = {gr-qc},
  url          = {https://arxiv.org/abs/2303.18046}
}

@article{SEOBNRv5EHM,
  author       = {Gamboa, Andres and others},
  title        = {SEOBNRv5EHM: Effective-one-body Waveform Model for Eccentric, Spin-aligned Binary Black Holes with Higher-order Modes},
  journal      = {arXiv e-prints},
  pages        = {arXiv:2412.12823},
  year         = {2024},
  eprint       = {2412.12823},
  archivePrefix= {arXiv},
  primaryClass = {gr-qc},
  url          = {https://arxiv.org/abs/2412.12823}
}

@article{Hinderer_2017,
   title={Foundations of an effective-one-body model for coalescing binaries on eccentric orbits},
   volume={96},
   ISSN={2470-0029},
   url={http://dx.doi.org/10.1103/PhysRevD.96.104048},
   DOI={10.1103/physrevd.96.104048},
   number={10},
   journal={Physical Review D},
   publisher={American Physical Society (APS)},
   author={Hinderer, Tanja and Babak, Stanislav},
   year={2017},
   month=nov }

@article{Chiaramello_2020,
   title={Faithful analytical effective-one-body waveform model for spin-aligned, moderately eccentric, coalescing black hole binaries},
   volume={101},
   ISSN={2470-0029},
   url={http://dx.doi.org/10.1103/PhysRevD.101.101501},
   DOI={10.1103/physrevd.101.101501},
   number={10},
   journal={Physical Review D},
   publisher={American Physical Society (APS)},
   author={Chiaramello, Danilo and Nagar, Alessandro},
   year={2020},
   month=may }

@article{Schmidt2015,
  author       = {Schmidt, Patricia and Ohme, Frank and Hannam, Mark},
  title        = {Towards models of gravitational waveforms from generic binaries: A simple approximate mapping between precessing and nonprecessing inspiral signals},
  journal      = {Physical Review D},
  volume       = {91},
  number       = {2},
  pages        = {024043},
  year         = {2015},
  doi          = {10.1103/PhysRevD.91.024043},
  eprint       = {1408.1810},
  archivePrefix= {arXiv},
  primaryClass = {gr-qc}
}

@article{Hannam2014,
  author       = {Hannam, Mark and Schmidt, Patricia and Bohé, Alejandro and Haegel, Leif and Husa, Sascha and Ohme, Frank and Pratten, Geraint and Pürrer, Michael},
  title        = {Simple model of complete precessing black-hole-binary gravitational waveforms},
  journal      = {Physical Review Letters},
  volume       = {113},
  pages        = {151101},
  year         = {2014},
  doi          = {10.1103/PhysRevLett.113.151101},
  eprint       = {1308.3271},
  archivePrefix= {arXiv},
  primaryClass = {gr-qc}
}

@misc{angles2018,
      title={Generative networks as inverse problems with Scattering transforms}, 
      author={Tomás Angles and Stéphane Mallat},
      year={2018},
      eprint={1805.06621},
      archivePrefix={arXiv},
      primaryClass={cs.LG},
      url={https://arxiv.org/abs/1805.06621}, 
}

@article{Khalouei_2025,
doi = {10.1088/1475-7516/2025/10/028},
url = {https://doi.org/10.1088/1475-7516/2025/10/028},
year = {2025},
month = {oct},
publisher = {IOP Publishing},
volume = {2025},
number = {10},
pages = {028},
author = {Khalouei, Elahe and Sabiu, Cristiano G. and Lee, Hyung Mok and Gopakumar, A.},
title = {External attention transformer: A robust AI model for identifying initial eccentricity signatures in binary black hole events in simulated advanced LIGO data},
journal = {Journal of Cosmology and Astroparticle Physics}
}

@misc{cheng2021,
      title={How to quantify fields or textures? A guide to the scattering transform}, 
      author={Sihao Cheng and Brice Ménard},
      year={2021},
      eprint={2112.01288},
      archivePrefix={arXiv},
      primaryClass={astro-ph.IM},
      url={https://arxiv.org/abs/2112.01288}, 
}

@article{Cheng__2020,
   title={A new approach to observational cosmology using the scattering transform},
   volume={499},
   ISSN={1365-2966},
   url={http://dx.doi.org/10.1093/mnras/staa3165},
   DOI={10.1093/mnras/staa3165},
   number={4},
   journal={Monthly Notices of the Royal Astronomical Society},
   publisher={Oxford University Press (OUP)},
   author={Cheng, Sihao and Ting, Yuan-Sen and Ménard, Brice and Bruna, Joan},
   year={2020},
   month=oct, pages={5902–5914} }

@article{Shimabukuro_2025,
   title={Analyzing the 21-cm forest with wavelet scattering transform: Insight into non-Gaussian features of the 21-cm forest},
   volume={112},
   ISSN={2470-0029},
   url={http://dx.doi.org/10.1103/nxr4-14gb},
   DOI={10.1103/nxr4-14gb},
   number={6},
   journal={Physical Review D},
   publisher={American Physical Society (APS)},
   author={Shimabukuro, Hayato and Xu, Yidong and Shao, Yue},
   year={2025},
   month=sep }

@article{Valogiannis_2022,
   title={Towards an optimal estimation of cosmological parameters with the wavelet scattering transform},
   volume={105},
   ISSN={2470-0029},
   url={http://dx.doi.org/10.1103/PhysRevD.105.103534},
   DOI={10.1103/physrevd.105.103534},
   number={10},
   journal={Physical Review D},
   publisher={American Physical Society (APS)},
   author={Valogiannis, Georgios and Dvorkin, Cora},
   year={2022},
   month=may }

@article{Valogiannis_2024,
   title={Towards unveiling the large-scale nature of gravity with the wavelet scattering transform},
   volume={2024},
   ISSN={1475-7516},
   url={http://dx.doi.org/10.1088/1475-7516/2024/11/061},
   DOI={10.1088/1475-7516/2024/11/061},
   number={11},
   journal={Journal of Cosmology and Astroparticle Physics},
   publisher={IOP Publishing},
   author={Valogiannis, Georgios and Villaescusa-Navarro, Francisco and Baldi, Marco},
   year={2024},
   month=nov, pages={061} }

@misc{siahkoohi2023,
      title={Unearthing InSights into Mars: Unsupervised Source Separation with Limited Data}, 
      author={Ali Siahkoohi and Rudy Morel and Maarten V. de Hoop and Erwan Allys and Grégory Sainton and Taichi Kawamura},
      year={2023},
      eprint={2301.11981},
      archivePrefix={arXiv},
      primaryClass={cs.LG},
      url={https://arxiv.org/abs/2301.11981}, 
}

@misc{morel2023a,
      title={Scale Dependencies and Self-Similar Models with Wavelet Scattering Spectra}, 
      author={Rudy Morel and Gaspar Rochette and Roberto Leonarduzzi and Jean-Philippe Bouchaud and Stéphane Mallat},
      year={2023},
      eprint={2204.10177},
      archivePrefix={arXiv},
      primaryClass={physics.data-an},
      url={https://arxiv.org/abs/2204.10177}, 
}

@misc{morel2023b,
      title={Path Shadowing Monte-Carlo}, 
      author={Rudy Morel and Stéphane Mallat and Jean-Philippe Bouchaud},
      year={2023},
      eprint={2308.01486},
      archivePrefix={arXiv},
      primaryClass={q-fin.MF},
      url={https://arxiv.org/abs/2308.01486}, 
}

%----- END -----%
\end{document}